\documentclass[acmsmall,screen]{acmart}

\setcopyright{none}
\usepackage{xspace}
\usepackage{xcolor}
\usepackage{graphicx}
\usepackage[normalem]{ulem} % normalem do not replace \emph to underline
\usepackage{amsmath}
\usepackage{enumitem}
\usepackage{soul} % \hl for highlighting text excerpts

\definecolor{MainColor}{RGB}{38,101,168}
\definecolor{MainFill}{RGB}{225,238,251}
\definecolor{MainHl}{RGB}{196,224,250}
\definecolor{DomainColor}{RGB}{35,115,52}
\definecolor{DomainFill}{RGB}{222,240,224}
\definecolor{CompetitionColor}{RGB}{158,118,0}
\definecolor{CompetitionFill}{RGB}{255,242,184}
\definecolor{ImprovementColor}{RGB}{196,72,72}
\definecolor{ImprovementFill}{RGB}{252,225,225}
\definecolor{KaggleBlue}{RGB}{32,190,255}
\sethlcolor{MainHl}
\definecolor{Link1Color}{RGB}{38,101,168}
\definecolor{Link1Hl}{RGB}{204,227,250}
\definecolor{Link2Color}{RGB}{126,72,176}
\definecolor{Link2Hl}{RGB}{231,217,248}
\definecolor{Link3Color}{RGB}{204,102,0}
\definecolor{Link3Hl}{RGB}{255,224,190}

\usepackage{hyperref}
\hypersetup{linkcolor=black,citecolor=black,anchorcolor=black,filecolor=black,menucolor=black,runcolor=black,urlcolor=black,hidelinks}
\usepackage{breakurl}

\usepackage{booktabs}
\usepackage{colortbl}
\usepackage{multirow}
\usepackage{makecell}
\usepackage{ragged2e}

\usepackage{caption}
\usepackage{subcaption}
\usepackage{wrapfig} % for small wrap figures - if needed

\usepackage{listings}
\usepackage[most]{tcolorbox} % file-like boxes in the example figures

\usepackage{tikz}
\usetikzlibrary{arrows.meta}
\usetikzlibrary{calc}
\usetikzlibrary{shapes.geometric}
\usetikzlibrary{shapes.symbols}
\usetikzlibrary{shadows}
\usetikzlibrary{decorations.pathreplacing}
\usetikzlibrary{positioning}
\usetikzlibrary{fit}
\colorlet{iconc}{black!70}
\colorlet{ovbg}{white}

\tikzset{
  pics/ov-trophy/.style={code={
    \fill[iconc] (-0.2,0.26) -- (0.2,0.26) -- (0.2,0.1) arc[start angle=0,end angle=-180,radius=0.2] -- cycle;
    \draw[iconc, line width=0.9pt] (0.19,0.21) -- (0.27,0.21) -- (0.27,0.13) to[out=-90,in=20] (0.14,-0.02);
    \draw[iconc, line width=0.9pt] (-0.19,0.21) -- (-0.27,0.21) -- (-0.27,0.13) to[out=-90,in=160] (-0.14,-0.02);
    \fill[iconc] (-0.035,-0.08) rectangle (0.035,-0.2);
    \fill[iconc] (-0.15,-0.2) rectangle (0.15,-0.27);
  }},
  pics/ov-clusters/.style={code={
    \foreach \cx/\cy/\pct in {-0.14/0.12/100, 0.16/0.12/70, 0.01/-0.15/45}
      \foreach \dx/\dy in {-0.06/0.02, 0.05/0.05, 0.01/-0.06}
        \fill[iconc!\pct] (\cx+\dx,\cy+\dy) circle (0.045);
  }},
  pics/ov-timeline/.style={code={
    \draw[iconc, line width=0.9pt, -{Latex[length=1.2mm]}] (-0.3,0) -- (0.33,0);
    \foreach \x in {-0.22,-0.1,0.02} \fill[iconc!55] (\x,0) circle (0.045);
    \draw[iconc, densely dashed, line width=0.5pt] (0.1,-0.2) -- (0.1,0.2);
    \fill[iconc] (0.19,0) circle (0.065);
  }},
  pics/ov-shield/.style={code={
    \fill[iconc] (0,0.27) .. controls (0.1,0.2) .. (0.23,0.2) -- (0.23,0.02)
      .. controls (0.23,-0.14) .. (0,-0.27) .. controls (-0.23,-0.14) .. (-0.23,0.02)
      -- (-0.23,0.2) .. controls (-0.1,0.2) .. cycle;
    \draw[white, line width=1.1pt, line cap=round, line join=round] (-0.1,0.0) -- (-0.02,-0.08) -- (0.11,0.09);
  }},
  pics/ov-lock/.style={code={
    \draw[iconc, line width=1.4pt] (-0.12,0.02) -- (-0.12,0.13) arc[start angle=180,end angle=0,radius=0.12] -- (0.12,0.02);
    \fill[iconc, rounded corners=1pt] (-0.21,-0.26) rectangle (0.21,0.04);
    \fill[white] (0,-0.08) circle (0.04);
    \fill[white] (-0.018,-0.08) rectangle (0.018,-0.18);
  }},
  pics/ov-robot/.style={code={
    \draw[iconc, line width=0.9pt] (0,0.18) -- (0,0.27);
    \fill[iconc] (0,0.29) circle (0.04);
    \fill[iconc] (-0.3,-0.07) rectangle (-0.24,0.05);
    \fill[iconc] (0.24,-0.07) rectangle (0.3,0.05);
    \fill[iconc, rounded corners=2.5pt] (-0.25,-0.22) rectangle (0.25,0.18);
    \fill[white] (-0.1,0.02) circle (0.05);
    \fill[white] (0.1,0.02) circle (0.05);
    \draw[white, line width=0.9pt, line cap=round] (-0.09,-0.11) -- (0.09,-0.11);
  }},
  pics/ov-layers/.style={code={
    \foreach \y/\pct in {-0.13/40, 0/70, 0.13/100}
      \filldraw[fill=iconc!\pct, draw=ovbg, line width=0.6pt] (0,\y+0.11) -- (0.28,\y) -- (0,\y-0.11) -- (-0.28,\y) -- cycle;
  }},
  pics/ov-medal/.style={code={
    \fill[iconc!55] (-0.15,0.28) -- (-0.04,0.28) -- (0.05,0.04) -- (-0.06,0.04) -- cycle;
    \fill[iconc!55] (0.15,0.28) -- (0.04,0.28) -- (-0.05,0.04) -- (0.06,0.04) -- cycle;
    \fill[iconc] (0,-0.1) circle (0.17);
    \draw[white, line width=0.7pt] (0,-0.1) circle (0.1);
  }},
  pics/ov-gear/.style={code={
    \foreach \a in {0,45,...,315} \fill[iconc, rotate=\a] (-0.055,0.14) rectangle (0.055,0.27);
    \fill[iconc] (0,0) circle (0.19);
    \fill[ovbg] (0,0) circle (0.075);
  }},
  pics/ov-book/.style={code={
    \fill[iconc] (-0.02,-0.2) -- (-0.3,-0.14) -- (-0.3,0.22) -- (-0.02,0.16) -- cycle;
    \fill[iconc!65] (0.02,-0.2) -- (0.3,-0.14) -- (0.3,0.22) -- (0.02,0.16) -- cycle;
    \foreach \y in {0.08,0.0,-0.08} {
      \draw[white, line width=0.5pt] (-0.25,\y+0.035) -- (-0.07,\y-0.005);
      \draw[white, line width=0.5pt] (0.07,\y-0.005) -- (0.25,\y+0.035);
    }
  }},
  pics/ov-terminal/.style={code={
    \fill[iconc, rounded corners=1.5pt] (-0.3,-0.22) rectangle (0.3,0.22);
    \fill[ovbg] (-0.26,-0.18) rectangle (0.26,0.12);
    \foreach \x in {-0.22,-0.16,-0.10} \fill[ovbg] (\x,0.17) circle (0.02);
    \draw[iconc, line width=0.9pt, line cap=round, line join=round] (-0.19,0.04) -- (-0.1,-0.02) -- (-0.19,-0.08);
    \draw[iconc, line width=0.9pt, line cap=round] (-0.04,-0.1) -- (0.1,-0.1);
  }},
  pics/ov-sliders/.style={code={
    \foreach \y/\x in {0.16/-0.1, 0/0.12, -0.16/-0.02} {
      \draw[iconc!60, line width=1pt, line cap=round] (-0.28,\y) -- (0.28,\y);
      \filldraw[fill=iconc, draw=ovbg, line width=0.6pt] (\x,\y) circle (0.06);
    }
  }},
  pics/ov-clock/.style={code={
    \draw[iconc, line width=1.4pt] (0,0) circle (0.24);
    \draw[iconc, line width=1pt, line cap=round] (0,0) -- (0,0.15);
    \draw[iconc, line width=1pt, line cap=round] (0,0) -- (0.11,-0.05);
    \fill[iconc] (0,0) circle (0.03);
  }},
  pics/ov-chart/.style={code={
    \fill[iconc!45] (-0.2,-0.24) rectangle (-0.08,-0.04);
    \fill[iconc] (-0.02,-0.24) rectangle (0.1,0.2);
    \fill[iconc!70] (0.16,-0.24) rectangle (0.28,0.06);
    \draw[iconc, line width=0.9pt] (-0.28,0.27) -- (-0.28,-0.24) -- (0.32,-0.24);
  }},
  pics/ov-magnifier/.style={code={
    \draw[iconc, line width=2.2pt, line cap=round] (0.08,-0.08) -- (0.25,-0.25);
    \draw[iconc, line width=1.4pt, fill=ovbg] (-0.06,0.06) circle (0.17);
    \node[text=iconc, font=\sffamily\bfseries\footnotesize, inner sep=0pt] at (-0.06,0.06) {?};
  }},
  pics/ov-tree/.style={code={
    \draw[iconc!70, line width=0.8pt] (-0.04,0.18) -- (-0.21,-0.02) (-0.04,0.18) -- (0.13,-0.02)
      (0.13,-0.02) -- (0.04,-0.22) (0.13,-0.02) -- (0.22,-0.22);
    \foreach \p in {(-0.04,0.18),(-0.21,-0.02),(0.13,-0.02),(0.04,-0.22),(0.22,-0.22)}
      \fill[iconc] \p circle (0.055);
  }},
  pics/ov-server/.style={code={
    \foreach \y in {0.13,-0.01,-0.15} {
      \fill[iconc, rounded corners=1pt] (-0.26,\y-0.055) rectangle (0.26,\y+0.055);
      \fill[ovbg] (0.15,\y) circle (0.022);
      \draw[ovbg, line width=0.5pt] (-0.19,\y) -- (0.0,\y);
    }
  }},
}

\usepackage{datetime} % for getting current date and time

\makeatletter
\newcommand{\DefMacro}{\@ifstar\@DefMacroAllowRedefine\@DefMacro}
\newcommand{\@DefMacro}[2]{\expandafter\newcommand\csname rmk-#1\endcsname{#2}}
\newcommand{\@DefMacroAllowRedefine}[2]{\expandafter\providecommand\csname rmk-#1\endcsname{} \expandafter\renewcommand\csname rmk-#1\endcsname{#2}}
\makeatother
\newcommand{\UseMacro}[1]{\csname rmk-#1\endcsname}

\newcommand{\XSpace}[1]{}
\newcommand{\XComment}[1]{}

\newcommand{\MyPara}[1]{\noindent\textbf{#1}.}

\newsavebox{\FitTableBox}
\newcommand{\FitTable}[1]{\sbox{\FitTableBox}{#1}%
  \ifdim\wd\FitTableBox>\linewidth\resizebox{\linewidth}{!}{\usebox{\FitTableBox}}\else\usebox{\FitTableBox}\fi}

\newtcolorbox{RQAnswer}[1]{enhanced, colback=MainFill!60, colframe=MainColor, boxrule=0.5pt,
  leftrule=2.5pt, arc=2pt, left=4pt, right=4pt, top=1pt, bottom=2pt, before skip=6pt, after skip=6pt,
  before upper={\noindent{\bfseries\color{MainColor}Answer to #1:}\ }}

\newcommand{\Code}[1]{\texttt{#1}}

\tcbset{
  filebox/.style={skin=bicolor, colback=black!5, colbacklower=white, colframe=black!45,
    boxrule=0.5pt, arc=1.5pt, boxsep=0pt, left=3pt, right=3pt, top=2pt, bottom=2pt, middle=2pt,
    toptitle=1.5pt, bottomtitle=1.5pt, fonttitle=\sffamily\scriptsize, fontupper=\scriptsize, fontlower=\scriptsize,
    nobeforeafter, box align=top,
    before upper={\RaggedRight}, before lower={\RaggedRight}},
  filebox type/.style={colframe=#1Color!70, colbacktitle=#1Fill, coltitle=#1Color!80!black},
  codebox/.style={enhanced, colback=white, colframe=black!45, colbacktitle=black!8, coltitle=black,
    boxrule=0.5pt, arc=1.5pt, boxsep=0pt, left=0pt, right=0pt, top=1.5pt, bottom=1.5pt,
    toptitle=1.5pt, bottomtitle=1.5pt, fonttitle=\sffamily\scriptsize, fontupper=\scriptsize\ttfamily,
    nobeforeafter, box align=top},
  webbox/.style={enhanced, colback=white, colframe=black!45, colbacktitle=black!8, coltitle=black,
    boxrule=0.5pt, arc=1.5pt, boxsep=0pt, left=3pt, right=3pt, top=2pt, bottom=2pt,
    toptitle=1.5pt, bottomtitle=1.5pt, fonttitle=\sffamily\scriptsize, fontupper=\scriptsize,
    nobeforeafter, box align=top, before upper={\RaggedRight}},
}
\newcommand{\WebURL}[1]{\begin{tcolorbox}[enhanced, colback=black!6, colframe=black!20, boxrule=0.3pt,
  arc=4pt, boxsep=0pt, left=4pt, right=4pt, top=1.5pt, bottom=1.5pt, nobeforeafter,
  fontupper=\sffamily\tiny\color{black!70}]#1\end{tcolorbox}\par\smallskip}
\newcommand{\WebTitle}[2]{{\sffamily\bfseries\fontsize{7.8}{9.5}\selectfont\hyphenpenalty=10000 #1\par}\vspace{1pt}{\color{black!60}#2\par}}
\newcommand{\WebSec}[1]{\par\smallskip{\sffamily\bfseries\fontsize{7.5}{9}\selectfont #1\strut\par}\nointerlineskip\vspace{0.5pt}%
  \noindent{\color{KaggleBlue}\rule{1.6em}{0.8pt}}\par\nointerlineskip\vspace{2pt}}
\newcommand{\FMKey}[1]{\texttt{\textbf{#1:}}~}
\newcommand{\KVLine}[2]{\par\hangindent=1.5em\hangafter=1\noindent\texttt{#1}:~#2}
\newcommand{\MDHead}[1]{\texttt{\textbf{\#\# #1}}\par}
\newcommand{\MDItem}{\makebox[1.6ex][l]{\texttt{-}}}
\newcommand{\LinkMark}[1]{\tikz[baseline=(m.base)]\node[circle, fill=Link#1Color, text=white, inner sep=0pt,
  minimum size=7pt, font=\sffamily\bfseries\fontsize{5.5}{5.5}\selectfont] (m) {#1};}
\newcommand{\CircledNum}[1]{\tikz[baseline=(m.base)]\node[circle, draw, line width=0.4pt, inner sep=0pt,
  minimum size=8pt, font=\fontsize{6.5}{6.5}\selectfont] (m) {#1};}
\newcommand{\LinkHl}[2]{{\sethlcolor{Link#1Hl}\LinkMark{#1}\,\hl{#2}}}
\newcommand{\CodeGutter}{3.2ex}
\newcommand{\CodeLn}[3]{\par\noindent{\setlength{\fboxsep}{0pt}%
  \colorbox{#1}{\makebox[\linewidth][l]{\strut\makebox[\CodeGutter][r]{#2}\hspace{0.8ex}#3}}}}
\newcommand{\CodeCtx}[1]{\CodeLn{white}{}{#1}}
\makeatletter
\newcommand{\CodeDel}{\@ifstar{\@CodeDiff{\LinkMark}{\textminus}{!55}}{\@CodeDiff{\@gobble}{\textminus}{!55}}}
\newcommand{\CodeAdd}{\@ifstar{\@CodeDiff{\LinkMark}{+}{}}{\@CodeDiff{\@gobble}{+}{}}}
\newcommand{\@CodeDiff}[5]{\CodeLn{Link#4Hl#3}{#1{#4}\,#2}{#5}}
\makeatother
\newcommand{\CodeGap}{\CodeLn{black!5}{}{}}
\newcommand{\CodeHunk}[1]{\CodeLn{black!8}{}{\color{black!60}@@ #1 @@}}
\newcommand{\CodeCm}[1]{{\color{black!55}\##1}}
\newcommand{\CodeStage}[1]{{\color{MainColor}\bfseries\##1}}
\newcommand{\CodeKw}[1]{\textbf{#1}}

\newcolumntype{R}[1]{>{\RaggedLeft\arraybackslash}p{#1}}
\newcolumntype{L}[1]{>{\RaggedRight\arraybackslash}p{#1}}

\definecolor{gray}{RGB}{211,211,211}
\newcommand{\jbasicstyle}{\small\sffamily} % Style of code

\newcommand{\jnumberstyle}{\scriptsize}

\lstdefinelanguage{pseudo}
{
  morekeywords={},
  keywordstyle=\bfseries,
  lineskip=-0.1em,
  numbers=left, % none for no numbers
  numberstyle=\jnumberstyle,
  numbersep=4pt,
  basicstyle=\jbasicstyle,
  breaklines=true,
  breakautoindent=true,
  tabsize=2,
  columns=fullflexible,
  morecomment=*[l][\textsl]{//},
  mathescape=true,
  xleftmargin=10pt,
}

\lstdefinelanguage{todo-comment}
{
  morekeywords={},
  keywordstyle=\bfseries,
  lineskip=-0.1em,
  numbers=none,
  basicstyle=\jbasicstyle,
  breaklines=true,
  breakautoindent=true,
  tabsize=2,
  columns=fullflexible,
  morecomment=*[l][\textsl]{//},
  mathescape=true,
  xleftmargin=-10pt,
}

\lstdefinelanguage{java-pretty}
{
  language=java,
  numbers=left,
  basicstyle=\scriptsize\ttfamily,
  numberstyle=\scriptsize,
  breaklines=true,
  columns=fullflexible,
  xleftmargin=16pt,
  showstringspaces=false,
}

\newcommand{\Tool}{\textsc{MLE-InsightForge}\xspace} % MLE-Assembler
\newcommand{\Benchmark}{\textsc{MLE-InsightBench}\xspace} % MLE-Insights-Bench
\newcommand{\Title}{Assembling Insights for Agentic Machine Learning Engineering Systems}
\newcommand{\NonDescendant}{causal non-descendant\xspace}
\newcommand{\NonDescendants}{causal non-descendants\xspace}
\newcommand{\NonDescendantSplit}{non-descendant split\xspace}
\newcommand{\TargetTask}{target task\xspace}
\newcommand{\TargetTasks}{target tasks\xspace}
\newcommand{\TargetTasksCap}{Target tasks\xspace}
\newcommand{\SourceCompetition}{source competition\xspace}
\newcommand{\SourceCompetitions}{source competitions\xspace}

\newcommand{\MLE}{MLE\xspace}
\newcommand{\MLECap}{Machine learning engineering\xspace}

\newcommand{\AgenticMLECap}{Agentic MLE\xspace}
\newcommand{\AgenticMLESystem}{agentic MLE system\xspace}
\newcommand{\AgenticMLESystems}{agentic MLE systems\xspace}
\newcommand{\AgenticMLESystemsCap}{Agentic MLE systems\xspace}
\newcommand{\MLEBench}{MLE-bench\xspace}
\newcommand{\MetaKaggle}{Meta Kaggle\xspace}

\newcommand{\CompetitionInsight}{competition insight\xspace}
\newcommand{\CompetitionInsights}{competition insights\xspace}
\newcommand{\CompetitionInsightCap}{Competition insight\xspace}
\newcommand{\CompetitionInsightsCap}{Competition insights\xspace}
\newcommand{\DomainInsight}{domain insight\xspace}
\newcommand{\DomainInsights}{domain insights\xspace}
\newcommand{\DomainInsightCap}{Domain insight\xspace}
\newcommand{\DomainInsightsCap}{Domain insights\xspace}
\newcommand{\ImprovementInsight}{improvement insight\xspace}
\newcommand{\ImprovementInsights}{improvement insights\xspace}
\newcommand{\ImprovementInsightCap}{Improvement insight\xspace}
\newcommand{\ImprovementInsightsCap}{Improvement insights\xspace}
\newcommand{\RawImprovementInsights}{raw improvement insights\xspace}

\newcommand{\InsightItems}{insight items\xspace}
\newcommand{\TriggerAction}{trigger--action\xspace}

\newcommand{\Aspects}{aspects\xspace}
\newcommand{\ExampleTask}{Energy Behavior of Prosumers\xspace}
\newcommand{\Skill}{skill\xspace}
\newcommand{\Skills}{skills\xspace}
\newcommand{\Playbook}{playbook\xspace}
\newcommand{\Playbooks}{playbooks\xspace}

\newcommand{\SkillModeCap}{Skill mode\xspace}
\newcommand{\PlaybookMode}{playbook mode\xspace}
\newcommand{\PlaybookModeCap}{Playbook mode\xspace}
\newcommand{\Baseline}{Baseline\xspace}
\newcommand{\FullSkill}{Full-skill\xspace}
\newcommand{\FullPlaybook}{Full-playbook\xspace}
\newcommand{\DomainPlaybook}{D-playbook\xspace}
\newcommand{\CompetitionPlaybook}{C-playbook\xspace}
\newcommand{\ImprovementPlaybook}{I-playbook\xspace}
\newcommand{\DCSkill}{DC-skill\xspace}
\newcommand{\DeepSeekModel}{\Code{deepseek-v4-flash-0731}\xspace}
\newcommand{\GPTModel}{\Code{GPT-6.1-Sol}\xspace}
\newcommand{\AssemblyModel}{\Code{gpt-5.6-sol-pro}\xspace}
\newcommand{\EmbeddingModel}{\Code{harrier-oss-v1-27b}\xspace}

\DefMacro{TH-Competition}{Competition\xspace}
\DefMacro{TH-Domain}{Domain\xspace}
\DefMacro{TH-TargetTask}{Target task\xspace}
\DefMacro{TH-Type}{Type\xspace}
\DefMacro{TH-Mode}{Mode\xspace}
\DefMacro{TH-DomainInsight}{Domain\xspace}
\DefMacro{TH-CompetitionInsight}{Competition\xspace}
\DefMacro{TH-ImprovementInsight}{Improvement\xspace}
\DefMacro{TH-Setting}{Setting\xspace}
\DefMacro{TH-Value}{Value\xspace}
\DefMacro{TH-Metric}{Metric\xspace}
\DefMacro{TH-LLM}{LLM\xspace}
\DefMacro{TH-Baseline}{\Baseline}
\DefMacro{TH-Skill}{\FullSkill}
\DefMacro{TH-Playbook}{\FullPlaybook}
\DefMacro{TH-DomainPlaybook}{\DomainPlaybook}
\DefMacro{TH-CompetitionPlaybook}{\CompetitionPlaybook}
\DefMacro{TH-ImprovementPlaybook}{\ImprovementPlaybook}
\DefMacro{TH-DCSkill}{\DCSkill}
\DefMacro{TH-ChangeFromBaseline}{Change from \Baseline}
\DefMacro{TH-PerTaskTime}{Per-task time [min]\xspace}
\DefMacro{TH-TotalTime}{Total time [day]\xspace}
\DefMacro{TH-PerTaskCost}{Per-task cost\xspace}
\DefMacro{TH-TotalCost}{Total cost\xspace}

\DefMacro{N-MinedCompetitions}{191\xspace}
\DefMacro{N-NonMLCompetitions}{18\xspace}
\DefMacro{N-ClusteredCompetitions}{173\xspace}
\DefMacro{N-DescriptionClusters}{12\xspace}
\DefMacro{N-ClusterOutliers}{7\xspace}
\DefMacro{N-ReassignedOutliers}{1\xspace}
\DefMacro{N-AudioCompetitions}{8\xspace}
\DefMacro{N-NoWriteupCompetitions}{4\xspace}
\DefMacro{N-RuntimeThresholdMin}{25.5\xspace}
\DefMacro{N-ImprovementClusters}{15\xspace}
\DefMacro{N-ImprovementOutlierShare}{4.9\%\xspace}
\DefMacro{N-CorpusCompetitions}{148\xspace}
\DefMacro{N-Domains}{12\xspace}
\DefMacro{N-EvalCompetitions}{12\xspace}
\DefMacro{N-Writeups}{1,308\xspace}
\DefMacro{N-GoldenSolutions}{2,309\xspace}
\DefMacro{N-CompetitionInsights}{148\xspace}
\DefMacro{N-ImprovementInsights}{1,068\xspace}
\DefMacro{N-SynthesizedImprovements}{16\xspace}
\DefMacro{N-BenchmarkCompetitions}{160\xspace}
\DefMacro{N-SameYearSources}{13\xspace}
\DefMacro{N-LaterSources}{8\xspace}
\DefMacro{N-LaterSourceDomains}{3\xspace}
\DefMacro{N-BlacklistEntries}{1,985\xspace}

\DefMacro{N-EvalSimple}{9\xspace}
\DefMacro{N-EvalCode}{3\xspace}

\DefMacro{N-BaselineRuns}{72\xspace}
\DefMacro{N-BaselineScore}{0.302\xspace}
\DefMacro{N-BaselineValidationScore}{0.468\xspace}
\DefMacro{N-BaselineInputTokens}{24.0M\xspace}
\DefMacro{N-BaselineReasoningTokens}{20.4K\xspace}
\DefMacro{N-BaselineOutputTokens}{114.9K\xspace}
\DefMacro{N-BaselineTimeMin}{228\xspace}
\DefMacro{N-BaselineSteps}{201\xspace}
\DefMacro{N-BaselineNodes}{6.4\xspace}
\DefMacro{N-BaselineGPUDay}{11.4\xspace}
\DefMacro{N-BaselineCostAvg}{\$1.72\xspace}
\DefMacro{N-BaselineCostTotal}{\$124\xspace}
\DefMacro{N-SkillRuns}{72\xspace}
\DefMacro{N-SkillScore}{0.676\xspace}
\DefMacro{N-SkillGain}{123.9\%\xspace}
\DefMacro{N-SkillValidationScore}{0.575\xspace}
\DefMacro{N-SkillInputTokens}{25.9M\xspace}
\DefMacro{N-SkillInputTokensChange}{7.7\%\xspace}
\DefMacro{N-SkillReasoningTokens}{68.1K\xspace}
\DefMacro{N-SkillReasoningTokensChange}{233.2\%\xspace}
\DefMacro{N-SkillOutputTokens}{123.3K\xspace}
\DefMacro{N-SkillOutputTokensChange}{7.2\%\xspace}
\DefMacro{N-SkillTimeMin}{292\xspace}
\DefMacro{N-SkillTimeMinChange}{28.0\%\xspace}
\DefMacro{N-SkillSteps}{191\xspace}
\DefMacro{N-SkillStepsChange}{-5.0\%\xspace}
\DefMacro{N-SkillNodes}{4.8\xspace}
\DefMacro{N-SkillNodesChange}{-24.3\%\xspace}
\DefMacro{N-SkillGPUDay}{14.6\xspace}
\DefMacro{N-SkillCostAvg}{\$1.88\xspace}
\DefMacro{N-SkillCostTotal}{\$135\xspace}
\DefMacro{N-PlaybookRuns}{72\xspace}
\DefMacro{N-PlaybookScore}{0.673\xspace}
\DefMacro{N-PlaybookGain}{122.8\%\xspace}
\DefMacro{N-PlaybookValidationScore}{0.548\xspace}
\DefMacro{N-PlaybookInputTokens}{30.1M\xspace}
\DefMacro{N-PlaybookInputTokensChange}{25.2\%\xspace}
\DefMacro{N-PlaybookReasoningTokens}{66.6K\xspace}
\DefMacro{N-PlaybookReasoningTokensChange}{225.9\%\xspace}
\DefMacro{N-PlaybookOutputTokens}{129.1K\xspace}
\DefMacro{N-PlaybookOutputTokensChange}{12.3\%\xspace}
\DefMacro{N-PlaybookTimeMin}{297\xspace}
\DefMacro{N-PlaybookTimeMinChange}{30.1\%\xspace}
\DefMacro{N-PlaybookSteps}{187\xspace}
\DefMacro{N-PlaybookStepsChange}{-7.1\%\xspace}
\DefMacro{N-PlaybookNodes}{5.9\xspace}
\DefMacro{N-PlaybookNodesChange}{-8.2\%\xspace}
\DefMacro{N-PlaybookGPUDay}{14.9\xspace}
\DefMacro{N-PlaybookCostAvg}{\$1.95\xspace}
\DefMacro{N-PlaybookCostTotal}{\$141\xspace}
\DefMacro{N-DomainPlaybookRuns}{72\xspace}
\DefMacro{N-DomainPlaybookScore}{0.595\xspace}
\DefMacro{N-DomainPlaybookGain}{96.9\%\xspace}
\DefMacro{N-DomainPlaybookValidationScore}{0.560\xspace}
\DefMacro{N-DomainPlaybookInputTokens}{27.3M\xspace}
\DefMacro{N-DomainPlaybookInputTokensChange}{13.6\%\xspace}
\DefMacro{N-DomainPlaybookReasoningTokens}{77.9K\xspace}
\DefMacro{N-DomainPlaybookReasoningTokensChange}{281.1\%\xspace}
\DefMacro{N-DomainPlaybookOutputTokens}{250.1K\xspace}
\DefMacro{N-DomainPlaybookOutputTokensChange}{117.6\%\xspace}
\DefMacro{N-DomainPlaybookTimeMin}{335\xspace}
\DefMacro{N-DomainPlaybookTimeMinChange}{46.6\%\xspace}
\DefMacro{N-DomainPlaybookSteps}{184\xspace}
\DefMacro{N-DomainPlaybookStepsChange}{-8.6\%\xspace}
\DefMacro{N-DomainPlaybookNodes}{5.6\xspace}
\DefMacro{N-DomainPlaybookNodesChange}{-13.0\%\xspace}
\DefMacro{N-DomainPlaybookGPUDay}{16.7\xspace}
\DefMacro{N-DomainPlaybookCostAvg}{\$2.00\xspace}
\DefMacro{N-DomainPlaybookCostTotal}{\$144\xspace}
\DefMacro{N-CompetitionPlaybookRuns}{72\xspace}
\DefMacro{N-CompetitionPlaybookScore}{0.608\xspace}
\DefMacro{N-CompetitionPlaybookGain}{101.4\%\xspace}
\DefMacro{N-CompetitionPlaybookValidationScore}{0.600\xspace}
\DefMacro{N-CompetitionPlaybookInputTokens}{25.4M\xspace}
\DefMacro{N-CompetitionPlaybookInputTokensChange}{5.4\%\xspace}
\DefMacro{N-CompetitionPlaybookReasoningTokens}{77.8K\xspace}
\DefMacro{N-CompetitionPlaybookReasoningTokensChange}{280.4\%\xspace}
\DefMacro{N-CompetitionPlaybookOutputTokens}{258.2K\xspace}
\DefMacro{N-CompetitionPlaybookOutputTokensChange}{124.6\%\xspace}
\DefMacro{N-CompetitionPlaybookTimeMin}{337\xspace}
\DefMacro{N-CompetitionPlaybookTimeMinChange}{47.8\%\xspace}
\DefMacro{N-CompetitionPlaybookSteps}{185\xspace}
\DefMacro{N-CompetitionPlaybookStepsChange}{-8.1\%\xspace}
\DefMacro{N-CompetitionPlaybookNodes}{6.1\xspace}
\DefMacro{N-CompetitionPlaybookNodesChange}{-5.0\%\xspace}
\DefMacro{N-CompetitionPlaybookGPUDay}{16.9\xspace}
\DefMacro{N-CompetitionPlaybookCostAvg}{\$1.74\xspace}
\DefMacro{N-CompetitionPlaybookCostTotal}{\$125\xspace}
\DefMacro{N-ImprovementPlaybookRuns}{72\xspace}
\DefMacro{N-ImprovementPlaybookScore}{0.611\xspace}
\DefMacro{N-ImprovementPlaybookGain}{102.1\%\xspace}
\DefMacro{N-ImprovementPlaybookValidationScore}{0.572\xspace}
\DefMacro{N-ImprovementPlaybookInputTokens}{26.7M\xspace}
\DefMacro{N-ImprovementPlaybookInputTokensChange}{10.9\%\xspace}
\DefMacro{N-ImprovementPlaybookReasoningTokens}{74.4K\xspace}
\DefMacro{N-ImprovementPlaybookReasoningTokensChange}{264.0\%\xspace}
\DefMacro{N-ImprovementPlaybookOutputTokens}{231.9K\xspace}
\DefMacro{N-ImprovementPlaybookOutputTokensChange}{101.8\%\xspace}
\DefMacro{N-ImprovementPlaybookTimeMin}{282\xspace}
\DefMacro{N-ImprovementPlaybookTimeMinChange}{23.6\%\xspace}
\DefMacro{N-ImprovementPlaybookSteps}{183\xspace}
\DefMacro{N-ImprovementPlaybookStepsChange}{-9.1\%\xspace}
\DefMacro{N-ImprovementPlaybookNodes}{5.9\xspace}
\DefMacro{N-ImprovementPlaybookNodesChange}{-7.4\%\xspace}
\DefMacro{N-ImprovementPlaybookGPUDay}{14.1\xspace}
\DefMacro{N-ImprovementPlaybookCostAvg}{\$1.77\xspace}
\DefMacro{N-ImprovementPlaybookCostTotal}{\$127\xspace}
\DefMacro{N-GPTBaselineRuns}{12\xspace}
\DefMacro{N-GPTBaselineScore}{0.346\xspace}
\DefMacro{N-GPTBaselineValidationScore}{0.518\xspace}
\DefMacro{N-GPTBaselineInputTokens}{34.8M\xspace}
\DefMacro{N-GPTBaselineReasoningTokens}{237.3K\xspace}
\DefMacro{N-GPTBaselineOutputTokens}{284.7K\xspace}
\DefMacro{N-GPTBaselineTimeMin}{436\xspace}
\DefMacro{N-GPTBaselineSteps}{145\xspace}
\DefMacro{N-GPTBaselineNodes}{6.4\xspace}
\DefMacro{N-GPTBaselineGPUDay}{3.6\xspace}
\DefMacro{N-GPTSkillRuns}{12\xspace}
\DefMacro{N-GPTSkillScore}{0.624\xspace}
\DefMacro{N-GPTSkillGain}{80.3\%\xspace}
\DefMacro{N-GPTSkillValidationScore}{0.489\xspace}
\DefMacro{N-GPTSkillInputTokens}{29.6M\xspace}
\DefMacro{N-GPTSkillInputTokensChange}{-14.9\%\xspace}
\DefMacro{N-GPTSkillReasoningTokens}{175.1K\xspace}
\DefMacro{N-GPTSkillReasoningTokensChange}{-26.2\%\xspace}
\DefMacro{N-GPTSkillOutputTokens}{223.3K\xspace}
\DefMacro{N-GPTSkillOutputTokensChange}{-21.6\%\xspace}
\DefMacro{N-GPTSkillTimeMin}{443\xspace}
\DefMacro{N-GPTSkillTimeMinChange}{1.7\%\xspace}
\DefMacro{N-GPTSkillSteps}{144\xspace}
\DefMacro{N-GPTSkillStepsChange}{-1.0\%\xspace}
\DefMacro{N-GPTSkillNodes}{7.2\xspace}
\DefMacro{N-GPTSkillNodesChange}{11.7\%\xspace}
\DefMacro{N-GPTSkillGPUDay}{3.7\xspace}
\DefMacro{N-GPTPlaybookRuns}{12\xspace}
\DefMacro{N-GPTPlaybookScore}{0.561\xspace}
\DefMacro{N-GPTPlaybookGain}{61.9\%\xspace}
\DefMacro{N-GPTPlaybookValidationScore}{0.526\xspace}
\DefMacro{N-GPTPlaybookInputTokens}{36.0M\xspace}
\DefMacro{N-GPTPlaybookInputTokensChange}{3.4\%\xspace}
\DefMacro{N-GPTPlaybookReasoningTokens}{206.7K\xspace}
\DefMacro{N-GPTPlaybookReasoningTokensChange}{-12.9\%\xspace}
\DefMacro{N-GPTPlaybookOutputTokens}{252.5K\xspace}
\DefMacro{N-GPTPlaybookOutputTokensChange}{-11.3\%\xspace}
\DefMacro{N-GPTPlaybookTimeMin}{467\xspace}
\DefMacro{N-GPTPlaybookTimeMinChange}{7.2\%\xspace}
\DefMacro{N-GPTPlaybookSteps}{146\xspace}
\DefMacro{N-GPTPlaybookStepsChange}{0.6\%\xspace}
\DefMacro{N-GPTPlaybookNodes}{8.8\xspace}
\DefMacro{N-GPTPlaybookNodesChange}{36.4\%\xspace}
\DefMacro{N-GPTPlaybookGPUDay}{3.9\xspace}
\DefMacro{N-BaselineTasks}{12\xspace}
\DefMacro{N-SkillTasks}{12\xspace}
\DefMacro{N-SkillTasksImproved}{12\xspace}
\DefMacro{N-SkillTasksSignificant}{11\xspace}
\DefMacro{N-SkillTasksSignificantHolm}{11\xspace}
\DefMacro{N-PlaybookTasks}{12\xspace}
\DefMacro{N-PlaybookTasksImproved}{11\xspace}
\DefMacro{N-PlaybookTasksSignificant}{11\xspace}
\DefMacro{N-PlaybookTasksSignificantHolm}{10\xspace}
\DefMacro{N-DomainPlaybookTasks}{12\xspace}
\DefMacro{N-DomainPlaybookTasksImproved}{12\xspace}
\DefMacro{N-DomainPlaybookTasksSignificant}{9\xspace}
\DefMacro{N-DomainPlaybookTasksSignificantHolm}{2\xspace}
\DefMacro{N-CompetitionPlaybookTasks}{12\xspace}
\DefMacro{N-CompetitionPlaybookTasksImproved}{11\xspace}
\DefMacro{N-CompetitionPlaybookTasksSignificant}{8\xspace}
\DefMacro{N-CompetitionPlaybookTasksSignificantHolm}{4\xspace}
\DefMacro{N-ImprovementPlaybookTasks}{12\xspace}
\DefMacro{N-ImprovementPlaybookTasksImproved}{11\xspace}
\DefMacro{N-ImprovementPlaybookTasksSignificant}{8\xspace}
\DefMacro{N-ImprovementPlaybookTasksSignificantHolm}{0\xspace}
\DefMacro{N-SkillAheadTasks}{6\xspace}
\DefMacro{N-PlaybookAheadTasks}{6\xspace}
\DefMacro{N-PlaybookAboveSingleTypesTasks}{7\xspace}
\DefMacro{N-SkillVsPlaybookTasksSignificant}{1\xspace}
\DefMacro{N-SkillVsPlaybookWilcoxonP}{0.79\xspace}
\DefMacro{N-ExampleBaselineMean}{228\xspace}
\DefMacro{N-ExampleBaselineMargin}{76.8\xspace}
\DefMacro{N-ExampleBaselineMin}{117\xspace}
\DefMacro{N-ExampleBaselineMax}{338\xspace}
\DefMacro{N-ExampleSkillMean}{64.3\xspace}
\DefMacro{N-ExampleSkillMargin}{2.84\xspace}
\DefMacro{N-ExampleSkillMin}{60.6\xspace}
\DefMacro{N-ExampleSkillMax}{67.7\xspace}
\DefMacro{N-ExamplePlaybookMean}{64.8\xspace}
\DefMacro{N-ExamplePlaybookMargin}{6.15\xspace}
\DefMacro{N-ExamplePlaybookMin}{57.1\xspace}
\DefMacro{N-ExamplePlaybookMax}{71.3\xspace}
\DefMacro{N-ExampleMedianThreshold}{69.1818\xspace}
\DefMacro{N-ExampleBronzeThreshold}{63.6529\xspace}
\DefMacro{N-ExampleRunBaselineValidation}{29.24\xspace}
\DefMacro{N-ExampleRunBaselinePrivate}{234.15\xspace}
\DefMacro{N-ExampleRunPlaybookValidation}{31.91\xspace}
\DefMacro{N-ExampleRunPlaybookPrivate}{57.12\xspace}
\DefMacro{N-IMCMedianThreshold}{29.26\xspace}
\DefMacro{N-IMCBronzeThreshold}{33.22\xspace}
\DefMacro{N-IMCSilverThreshold}{37.34\xspace}
\DefMacro{N-IMCGoldThreshold}{43.34\xspace}
\DefMacro{N-DeepSeekIMCBaselineScore}{21.1\xspace}
\DefMacro{N-DeepSeekIMCBaselineSteps}{213\xspace}
\DefMacro{N-DeepSeekIMCBaselineNodes}{4.0\xspace}
\DefMacro{N-DeepSeekIMCBaselineTimeMin}{245\xspace}
\DefMacro{N-DeepSeekIMCBaselineInspections}{9.8\xspace}
\DefMacro{N-DeepSeekIMCBaselineCostAvg}{\$2.91\xspace}
\DefMacro{N-DeepSeekIMCSkillScore}{35.2\xspace}
\DefMacro{N-DeepSeekIMCSkillGain}{680.8\%\xspace}
\DefMacro{N-DeepSeekIMCSkillSteps}{225\xspace}
\DefMacro{N-DeepSeekIMCSkillNodes}{2.5\xspace}
\DefMacro{N-DeepSeekIMCSkillTimeMin}{210\xspace}
\DefMacro{N-DeepSeekIMCSkillInspections}{8.2\xspace}
\DefMacro{N-DeepSeekIMCSkillCostAvg}{\$2.39\xspace}
\DefMacro{N-DeepSeekIMCPlaybookScore}{33.8\xspace}
\DefMacro{N-DeepSeekIMCPlaybookGain}{613.4\%\xspace}
\DefMacro{N-DeepSeekIMCPlaybookSteps}{216\xspace}
\DefMacro{N-DeepSeekIMCPlaybookNodes}{4.0\xspace}
\DefMacro{N-DeepSeekIMCPlaybookTimeMin}{320\xspace}
\DefMacro{N-DeepSeekIMCPlaybookInspections}{5.3\xspace}
\DefMacro{N-DeepSeekIMCPlaybookCostAvg}{\$2.57\xspace}
\DefMacro{N-GPTIMCBaselineScore}{42.8\xspace}
\DefMacro{N-GPTIMCBaselineSteps}{160\xspace}
\DefMacro{N-GPTIMCBaselineNodes}{9.2\xspace}
\DefMacro{N-GPTIMCBaselineTimeMin}{594\xspace}
\DefMacro{N-GPTIMCBaselineInspections}{3.3\xspace}
\DefMacro{N-GPTIMCSkillScore}{52.0\xspace}
\DefMacro{N-GPTIMCSkillGain}{427.9\%\xspace}
\DefMacro{N-GPTIMCSkillSteps}{135\xspace}
\DefMacro{N-GPTIMCSkillNodes}{11.5\xspace}
\DefMacro{N-GPTIMCSkillTimeMin}{629\xspace}
\DefMacro{N-GPTIMCSkillInspections}{3.7\xspace}
\DefMacro{N-GPTIMCPlaybookScore}{49.9\xspace}
\DefMacro{N-GPTIMCPlaybookGain}{326.9\%\xspace}
\DefMacro{N-GPTIMCPlaybookSteps}{134\xspace}
\DefMacro{N-GPTIMCPlaybookNodes}{13.3\xspace}
\DefMacro{N-GPTIMCPlaybookTimeMin}{621\xspace}
\DefMacro{N-GPTIMCPlaybookInspections}{3.5\xspace}
\DefMacro{N-POGMedianThreshold}{0.016845\xspace}
\DefMacro{N-POGBronzeThreshold}{0.01518\xspace}
\DefMacro{N-POGSilverThreshold}{0.01459\xspace}
\DefMacro{N-POGGoldThreshold}{0.01414\xspace}
\DefMacro{N-DeepSeekPOGBaselineScore}{0.0255\xspace}
\DefMacro{N-DeepSeekPOGBaselineSteps}{193\xspace}
\DefMacro{N-DeepSeekPOGBaselineNodes}{7.2\xspace}
\DefMacro{N-DeepSeekPOGBaselineTimeMin}{152\xspace}
\DefMacro{N-DeepSeekPOGBaselineInspections}{6.5\xspace}
\DefMacro{N-DeepSeekPOGBaselineCostAvg}{\$0.77\xspace}
\DefMacro{N-DeepSeekPOGSkillScore}{0.0249\xspace}
\DefMacro{N-DeepSeekPOGSkillGain}{7.2\%\xspace}
\DefMacro{N-DeepSeekPOGSkillSteps}{159\xspace}
\DefMacro{N-DeepSeekPOGSkillNodes}{5.8\xspace}
\DefMacro{N-DeepSeekPOGSkillTimeMin}{298\xspace}
\DefMacro{N-DeepSeekPOGSkillInspections}{5.3\xspace}
\DefMacro{N-DeepSeekPOGSkillCostAvg}{\$1.03\xspace}
\DefMacro{N-DeepSeekPOGPlaybookScore}{0.0273\xspace}
\DefMacro{N-DeepSeekPOGPlaybookGain}{-24.2\%\xspace}
\DefMacro{N-DeepSeekPOGPlaybookSteps}{158\xspace}
\DefMacro{N-DeepSeekPOGPlaybookNodes}{7.3\xspace}
\DefMacro{N-DeepSeekPOGPlaybookTimeMin}{286\xspace}
\DefMacro{N-DeepSeekPOGPlaybookInspections}{7.7\xspace}
\DefMacro{N-DeepSeekPOGPlaybookCostAvg}{\$1.15\xspace}
\DefMacro{N-GPTPOGBaselineScore}{0.0261\xspace}
\DefMacro{N-GPTPOGBaselineSteps}{130\xspace}
\DefMacro{N-GPTPOGBaselineNodes}{3.7\xspace}
\DefMacro{N-GPTPOGBaselineTimeMin}{277\xspace}
\DefMacro{N-GPTPOGBaselineInspections}{4.0\xspace}
\DefMacro{N-GPTPOGSkillScore}{0.0251\xspace}
\DefMacro{N-GPTPOGSkillGain}{38.2\%\xspace}
\DefMacro{N-GPTPOGSkillSteps}{152\xspace}
\DefMacro{N-GPTPOGSkillNodes}{2.8\xspace}
\DefMacro{N-GPTPOGSkillTimeMin}{257\xspace}
\DefMacro{N-GPTPOGSkillInspections}{4.0\xspace}
\DefMacro{N-GPTPOGPlaybookScore}{0.0253\xspace}
\DefMacro{N-GPTPOGPlaybookGain}{29.8\%\xspace}
\DefMacro{N-GPTPOGPlaybookSteps}{159\xspace}
\DefMacro{N-GPTPOGPlaybookNodes}{4.2\xspace}
\DefMacro{N-GPTPOGPlaybookTimeMin}{313\xspace}
\DefMacro{N-GPTPOGPlaybookInspections}{5.2\xspace}
\DefMacro{N-DeepSeekLLMTasksBaselineGPUDay}{1.7\xspace}
\DefMacro{N-DeepSeekLLMTasksBaselineCostTotal}{\$22\xspace}
\DefMacro{N-DeepSeekLLMTasksSkillGain}{61.8\%\xspace}
\DefMacro{N-DeepSeekLLMTasksSkillGPUDay}{2.1\xspace}
\DefMacro{N-DeepSeekLLMTasksSkillCostTotal}{\$21\xspace}
\DefMacro{N-DeepSeekLLMTasksPlaybookGain}{27.5\%\xspace}
\DefMacro{N-DeepSeekLLMTasksPlaybookGPUDay}{2.5\xspace}
\DefMacro{N-DeepSeekLLMTasksPlaybookCostTotal}{\$22\xspace}
\DefMacro{N-GPTLLMTasksBaselineGPUDay}{3.6\xspace}
\DefMacro{N-GPTLLMTasksBaselineCostTotal}{\$153\xspace}
\DefMacro{N-GPTLLMTasksSkillGain}{80.3\%\xspace}
\DefMacro{N-GPTLLMTasksSkillGPUDay}{3.7\xspace}
\DefMacro{N-GPTLLMTasksSkillCostTotal}{\$128\xspace}
\DefMacro{N-GPTLLMTasksPlaybookGain}{61.9\%\xspace}
\DefMacro{N-GPTLLMTasksPlaybookGPUDay}{3.9\xspace}
\DefMacro{N-GPTLLMTasksPlaybookCostTotal}{\$160\xspace}
\DefMacro{N-GPTProjectedCostMax}{\$957\xspace}

\DefMacro{N-CurveBaselineBestFirst}{52.1\xspace}
\DefMacro{N-CurveBaselineBestFinal}{55.3\xspace}
\DefMacro{N-CurveBaselineGainFirst}{-0.0\xspace}
\DefMacro{N-CurveBaselineGainFinal}{3.3\xspace}
\DefMacro{N-CurveSkillBestFirst}{53.9\xspace}
\DefMacro{N-CurveSkillBestFinal}{59.6\xspace}
\DefMacro{N-CurveSkillGainFirst}{1.8\xspace}
\DefMacro{N-CurveSkillGainFinal}{7.8\xspace}
\DefMacro{N-CurvePlaybookBestFirst}{55.0\xspace}
\DefMacro{N-CurvePlaybookBestFinal}{57.8\xspace}
\DefMacro{N-CurvePlaybookGainFirst}{2.9\xspace}
\DefMacro{N-CurvePlaybookGainFinal}{5.9\xspace}
\DefMacro{N-CurveSkillReachBaselineMin}{80\xspace}
\DefMacro{N-CurvePlaybookReachBaselineMin}{40\xspace}
\DefMacro{N-CurveAblationBaselineGainFinal}{3.9\xspace}
\DefMacro{N-CurveDomainPlaybookGainFinal}{7.1\xspace}
\DefMacro{N-CurveDomainPlaybookEffectFinal}{3.2\xspace}
\DefMacro{N-CurveDomainPlaybookEffectAtHundred}{3.0\xspace}
\DefMacro{N-CurveCompetitionPlaybookGainFinal}{6.7\xspace}
\DefMacro{N-CurveCompetitionPlaybookEffectFinal}{6.1\xspace}
\DefMacro{N-CurveCompetitionPlaybookEffectAtHundred}{2.2\xspace}
\DefMacro{N-CurveImprovementPlaybookGainFinal}{8.2\xspace}
\DefMacro{N-CurveImprovementPlaybookEffectFinal}{4.4\xspace}
\DefMacro{N-CurveImprovementPlaybookEffectAtHundred}{3.1\xspace}
\DefMacro{N-VennAllMain}{509,717\xspace}
\DefMacro{N-VennBaselineOnly}{3,838\xspace}
\DefMacro{N-VennSkillOnly}{5,376\xspace}
\DefMacro{N-VennPlaybookOnly}{4,918\xspace}
\DefMacro{N-VennSkillPlaybookOnly}{8,786\xspace}
\DefMacro{N-VennInsightsNotBaseline}{19,080\xspace}
\DefMacro{N-VennAblationBaselineOnly}{1,184\xspace}
\DefMacro{N-VennDomainPlaybookOnly}{1,238\xspace}
\DefMacro{N-VennCompetitionPlaybookOnly}{12,345\xspace}
\DefMacro{N-VennImprovementPlaybookOnly}{4,566\xspace}

\DefMacro{N-SkillReadsRuns}{72\xspace}
\DefMacro{N-SkillReadsReadRuns}{19\xspace}
\DefMacro{N-SkillReadsReadCompetitionRuns}{14\xspace}
\DefMacro{N-SkillReadsReadImprovementRuns}{5\xspace}
\DefMacro{N-SkillReadsReadDomainRuns}{2\xspace}
\DefMacro{N-GPTSkillReadsRuns}{12\xspace}
\DefMacro{N-GPTSkillReadsReadRuns}{12\xspace}
\DefMacro{N-GPTSkillReadsReadCompetitionRuns}{6\xspace}
\DefMacro{N-GPTSkillReadsReadImprovementRuns}{12\xspace}
\DefMacro{N-GPTSkillReadsReadDomainRuns}{6\xspace}
\DefMacro{N-AuditRuns}{468\xspace}
\DefMacro{N-AuditWebRuns}{64\xspace}
\DefMacro{N-AuditTargetRuns}{10\xspace}
\DefMacro{N-AuditTargetSolutionRuns}{6\xspace}
\DefMacro{N-AuditTargetSolutionDeepSeekRuns}{3\xspace}
\DefMacro{N-AuditTargetSolutionGPTRuns}{3\xspace}
\DefMacro{N-AuditTargetMetricRuns}{3\xspace}
\DefMacro{N-AuditTargetFailedRuns}{1\xspace}
\DefMacro{N-DeepSeekIMCFirstNodeStep}{165\xspace}
\DefMacro{N-DeepSeekIMCFirstNodeMin}{75\xspace}
\DefMacro{N-DeepSeekPOGFirstNodeStep}{44\xspace}
\DefMacro{N-DeepSeekPOGFirstNodeMin}{18\xspace}
\DefMacro{N-GPTIMCFirstNodeStep}{19\xspace}
\DefMacro{N-GPTIMCFirstNodeMin}{8\xspace}
\DefMacro{N-GPTPOGFirstNodeStep}{9\xspace}
\DefMacro{N-GPTPOGFirstNodeMin}{3\xspace}

\DefMacro{N-AssemblyRuns}{471\xspace}
\DefMacro{N-AssemblyCost}{903\xspace}
\DefMacro{N-AssemblyCostAll}{1,544\xspace}
\DefMacro{N-AssemblyCostCompetition}{565\xspace}
\DefMacro{N-AssemblyCostDomain}{22\xspace}
\DefMacro{N-AssemblyCostImprovement}{317\xspace}
\DefMacro{N-AssemblyInputTokensM}{977M\xspace}
\DefMacro{N-AssemblyOutputTokensM}{33.5M\xspace}

\renewcommand\footnotetextcopyrightpermission[1]{} % removes footnote with conference information in first column
\begin{document}

\title{\Title}
%\title[Short Title]{Full Title}         %% [Short Title] is optional;
                                        %% when present, will be used in
                                        %% header instead of Full Title.
%\titlenote{with title note}             %% \titlenote is optional;
                                        %% can be repeated if necessary;
                                        %% contents suppressed with 'anonymous'
%\subtitle{Subtitle}                     %% \subtitle is optional
%\subtitlenote{with subtitle note}       %% \subtitlenote is optional;
                                        %% can be repeated if necessary;
                                        %% contents suppressed with 'anonymous'

%% Author information
%% Contents and number of authors suppressed with 'anonymous'.
%% Each author should be introduced by \author, followed by
%% \authornote (optional), \orcid (optional), \affiliation, and
%% \email.
%% An author may have multiple affiliations and/or emails; repeat the
%% appropriate command.
%% Many elements are not rendered, but should be provided for metadata
%% extraction tools.

%% Author with single affiliation.
% \author{Bihui Jin, Yinxi Li, Kaiyuan Wang, Pengyu Nie}
%\authornote{with author1 note}          %% \authornote is optional;
                                        %% can be repeated if necessary
\author{Bihui Jin}
\orcid{0009-0009-0011-1134}
\affiliation{
  \institution{University of Waterloo}
  \city{Waterloo}
  \country{Canada}
}
\email{bihui.jin@uwaterloo.ca}

\author{Yinxi Li}
\orcid{0009-0006-4337-4848}
\affiliation{
  \institution{University of Waterloo}
  \city{Waterloo}
  \country{Canada}
}
\email{yinxi.li@uwaterloo.ca}

\author{Kaiyuan Wang}
\orcid{0000-0002-1790-0721}
\affiliation{
  \institution{Google Inc.}
  \city{Mountain View}
  \country{USA}
}
\email{kaiyuanw@google.com}

\author{Pengyu Nie}
\orcid{0000-0003-1529-3216}
\affiliation{
  \institution{University of Waterloo}
  \city{Waterloo}
  \country{Canada}
}
\email{pynie@uwaterloo.ca}

%% Author with two affiliations and emails.
%% \author{First2 Last2}
%% \authornote{with author2 note}          %% \authornote is optional;
%%                                         %% can be repeated if necessary
%% \orcid{nnnn-nnnn-nnnn-nnnn}             %% \orcid is optional
%% \affiliation{
%%   \position{Position2a}
%%   \department{Department2a}             %% \department is recommended
%%   \institution{Institution2a}           %% \institution is required
%%   \streetaddress{Street2a Address2a}
%%   \city{City2a}
%%   \state{State2a}
%%   \postcode{Post-Code2a}
%%   \country{Country2a}                   %% \country is recommended
%% }
%% \email{first2.last2@inst2a.com}         %% \email is recommended
%% \affiliation{
%%   \position{Position2b}
%%   \department{Department2b}             %% \department is recommended
%%   \institution{Institution2b}           %% \institution is required
%%   \streetaddress{Street3b Address2b}
%%   \city{City2b}
%%   \state{State2b}
%%   \postcode{Post-Code2b}
%%   \country{Country2b}                   %% \country is recommended
%% }
%% \email{first2.last2@inst2b.org}         %% \email is recommended

%% For adding cover page
%% \thispagestyle{empty}
%% \input{COVER}
%% \newpage
%% \pagenumbering{arabic}
%% \setcounter{page}{1}

%% Abstract
\begin{abstract}

Agentic machine learning engineering (MLE) is an emerging AI4SE application, targeting complex machine learning tasks and taking a step toward the recursive self-improvement of AI systems.
Recent agentic MLE systems highlight the importance of collecting and leveraging insights from closely-related MLE tasks:
some systems condition code generation on expert domain knowledge, which is implicitly curated from peer MLE tasks;
some systems have a loop of solving an MLE task, gathering validation feedback, and refining their prompts, skills, or memory to benefit future tasks.
However, collecting and injecting insights is ad hoc, and existing evaluations do not control the sources of insights, leaving data leakage a threat to validity.

To close these gaps, we systematically study how insights assembled from closely-related MLE tasks affect an agentic MLE system, and how their effect depends on the type of insights and the way they are injected.
We first collect a new benchmark, \Benchmark, consisting of \UseMacro{N-BenchmarkCompetitions} Kaggle competitions clustered into \UseMacro{N-Domains} domains.
\Benchmark is designed to keep \TargetTask information out of the insights:
one competition from each domain is held out as a \TargetTask, and only the remaining \UseMacro{N-CorpusCompetitions} competitions, none of which can have been influenced by the \TargetTasks (e.g., as a later edition of the same series or by reusing their solutions), serve as \SourceCompetitions.
We then design and implement an insight assembly agent, \Tool, that explores the \SourceCompetitions, their top human-written solutions, and writeups, and assembles them into three types of insights:
\UseMacro{N-Domains} \DomainInsights (summarizing domain-level best practices), \UseMacro{N-CompetitionInsights} \CompetitionInsights (describing competition-level winning solutions), and \UseMacro{N-SynthesizedImprovements} \ImprovementInsights (detailing generic debugging and improvement strategies).
The extracted insights can be injected into an agentic MLE system as skills, which the agent reads selectively on demand, or as a playbook placed upfront in its initial context.
In our evaluation, the extracted insights improve the normalized private-test score of an agentic MLE system by \UseMacro{N-SkillGain} over the baseline without insights.
Further controlled experiments show that:
the three types of insights are complementary, 
the better injection mode (skill vs. playbook) varies across tasks, with skills being more cost-effective overall,
and the improvements from insights generalize across backend LLMs.

\end{abstract}

%% Computing Classification Scheme
%% The code below should be generated by the tool at
%% http://dl.acm.org/ccs.cfm
%% \begin{CCSXML}
%%   <ccs2012>
%%   <concept>
%%   <concept_id>10011007.10011006.10011041.10011047</concept_id>
%%   <concept_desc>Software and its engineering~Source code generation</concept_desc>
%%   <concept_significance>300</concept_significance>
%%   </concept>
%%   </ccs2012>
%% \end{CCSXML}
%
%% \ccsdesc[300]{Software and its engineering~Source code generation}
%
%% End of generated code

%% \keywords{}

\maketitle

%% Change the authors list of running head, if it is too long to fit
%% \renewcommand{\shortauthors}{Nie et al.}

%% Contents
\section{Introduction}
\label{sec:intro}

\MLECap (MLE) is the end-to-end work of turning a dataset and a modeling objective into an executable machine learning (ML) solution, including data inspection, model selection, validation design, debugging, and compute allocation.
ML solutions are typically written in Python, as Jupyter notebooks or scripts~\cite{PimentelETAL19Notebooks,GrotovETAL22NotebooksScripts}, and the SE community has studied them as software artifacts with distinct engineering challenges, such as data management, code quality, and reproducibility~\cite{AmershiETAL19SE4ML,PimentelETAL19Notebooks,MostafaviETAL24DistilKaggle,JinETAL26Modernization}.
This makes ML solutions a natural target of LLM-based coding agents, but also a challenging one.
Existing code generation benchmarks, including those for data science code, check a generated function, notebook cell, or patch against tests or reference outputs~\cite{LaiETAL23DS1000,YinETAL23ARCADE,JimenezETAL24SWEbench}.
An ML solution, in contrast, has no test oracle: its quality is a metric value on unseen data, which can only be estimated by executing long-running training code and improved by repeatedly revising the solution.

\AgenticMLECap is an emerging application of coding agents in which LLMs perform this work by iteratively writing, executing, and refining code~\cite{HuangETAL24MLAgentBench,ChanETAL25MLEBench,JiangETAL25AIDE}.
\AgenticMLESystemsCap target complex ML tasks, such as Kaggle competitions, and are a step toward the recursive self-improvement of AI systems, since a system that can engineer ML solutions can, in principle, contribute to building better AI systems~\cite{ChanETAL25MLEBench}.
Evolution has been rapid: on \MLEBench~\cite{ChanETAL25MLEBench}, the fraction of competitions in which the best system earns a medal grew from 17\% in late 2024 to over 60\% in early 2026.

Recent \AgenticMLESystems highlight the importance of collecting and leveraging insights from closely-related \MLE tasks.
Some systems generate code based on expert domain knowledge, which is implicitly curated from peer \MLE tasks, either by searching the web for state-of-the-art models and winning solutions of similar tasks~\cite{NamETAL25MLESTAR,ChenETAL26MARS} or by drawing on knowledge bases synthesized from open-source repositories, competition platforms, and competition discussions~\cite{DuETAL26MLEvolve,DuETAL25AutoMLGen,NadafianETAL26KAPSO}.
Other systems have a loop of solving an \MLE task, gathering validation feedback, and refining their prompts, skills, or memory to benefit future tasks~\cite{GrosnitETAL24AgentK,ZhuETAL26MLMaster2,ZhangETAL26AIBuildAI2}.

However, in these systems, the process of collecting and injecting insights is ad hoc: each system fixes one design, without a controlled comparison of the alternatives.
Each system collects insights from different sources (web search, curated knowledge bases, or its own past runs), represents them differently, and injects them into the agent's context in its own way.
The ablation studies in these work typically remove the entire insight source, e.g., the knowledge base or the cache of prior wisdom~\cite{DuETAL25AutoMLGen,ZhuETAL26MLMaster2}, which shows that insights help as a whole but leaves open which types of insights contribute and how they should be injected.

Moreover, existing evaluations of these systems do not control for the sources of insights, which leaves data leakage as a threat to validity: the sources are chosen by each system rather than fixed by the benchmark, so whether they exclude the \TargetTasks is left to each system and rarely verifiable.
For example, web search at solving time~\cite{NamETAL25MLESTAR,ChenETAL26MARS} can retrieve the winning writeups of the very competition being evaluated, and knowledge bases built from competition platforms may contain its solutions; even when the benchmark competitions are excluded by name~\cite{ZhuETAL26MLMaster2,YangETAL25RDAgent}, closely-related editions of the same competition series can remain.
Such concerns were raised on the \MLEBench leaderboard~\cite{MLEBenchIssue138}, which has since paused new submissions to ensure fair and comparable results.

\begin{figure*}[t]
\centering
\resizebox{0.8\textwidth}{!}{\newcommand{\OvBlock}[4]{%
  \def\ovx{#1}%
  \colorlet{iconc}{#2}%
  \colorlet{ovbg}{#2!6}%
  \begin{scope}
    \clip[rounded corners=4pt] (#1,0) rectangle ++(4.7,-7.2);
    \fill[ovbg] (#1,0) rectangle ++(4.7,-7.2);
    \fill[iconc] (#1,0) rectangle ++(4.7,-1.0);
  \end{scope}
  \draw[iconc, rounded corners=4pt, line width=0.6pt] (#1,0) rectangle ++(4.7,-7.2);
  \node[text=white, align=center, inner sep=0pt] at ({#1+2.35},-0.5) {\textbf{\small #3}\\[1pt]\footnotesize #4};
}
\newcommand{\OvItem}[4]{%
  \pic at ({\ovx+0.45},{-1.55-1.2*(#1)}) {#2};
  \node[anchor=north west, align=left, text width=3.7cm, inner sep=0pt, font=\sffamily\scriptsize,
    execute at begin node={\hyphenpenalty=10000\exhyphenpenalty=10000}]
    at ({\ovx+0.88},{-1.2-1.2*(#1)}) {\textbf{\footnotesize #3}\\[0.5pt]#4};
}
\begin{tikzpicture}[x=1cm, y=1cm, font=\sffamily]

\OvBlock{0}{blue!55!black}{Benchmark}{\Benchmark (\S\ref*{sec:benchmark})}
\OvItem{0}{ov-trophy}{Kaggle competitions}{\UseMacro{N-BenchmarkCompetitions} high-quality competitions in 2020--2025}
\OvItem{1}{ov-clusters}{Domain clustering}{\UseMacro{N-Domains} domains of closely-related tasks}
\OvItem{2}{ov-timeline}{Held-out split}{target (\UseMacro{N-EvalCompetitions}): one per domain;\\source (\UseMacro{N-CorpusCompetitions}): \NonDescendants}
\OvItem{3}{ov-shield}{Data leakage control}{target materials blocked during insight assembly;\\no web search when solving}
\OvItem{4}{ov-lock}{Private-test grading}{\MLEBench-compatible graders; private-test set never exposed}

\OvBlock{4.9}{orange!75!black}{Technique}{\Tool (\S\ref*{sec:technique})}
\OvItem{0}{ov-robot}{Insight assembly agent}{reads all sources; records the evidence of each insight}
\OvItem{1}{ov-layers}{\UseMacro{N-Domains} domain insights}{domain-level best practices}
\OvItem{2}{ov-medal}{\UseMacro{N-CompetitionInsights} competition insights}{decisions of the winning solutions}
\OvItem{3}{ov-gear}{\UseMacro{N-SynthesizedImprovements} improvement insights}{consolidated debugging and improvement strategies}
\OvItem{4}{ov-book}{Insight injection}{as skills (on demand) or playbooks (in the initial context)}

\OvBlock{9.8}{green!40!black}{Evaluation}{Systematic study (\S\ref*{sec:eval})}
\OvItem{0}{ov-terminal}{Solving agent}{State-of-the-art ReAct agent developing a solution tree}
\OvItem{1}{ov-sliders}{Six conditions}{\Baseline, \FullSkill, \FullPlaybook, and three ablations}
\OvItem{2}{ov-clock}{Fixed budget}{12 hours, 250 steps, and one GPU per run}
\OvItem{3}{ov-chart}{Metrics}{normalized private-test score, token consumption, and time}
\OvItem{4}{ov-magnifier}{Ablation studies}{insight types, injection modes, and backend LLMs}

\end{tikzpicture}}
\vspace{-4pt}
\caption{Overview of our study.}\label{fig:overview}
\vspace{-8pt}
\Description{Three side-by-side panels. The benchmark panel lists Kaggle competition collection, domain clustering, the held-out split, leakage control, and private-test grading. The technique panel lists the insight assembly agent, domain, competition, and improvement insights, and insight injection as skills or playbooks. The evaluation panel lists the agentic MLE system, six conditions, the fixed budget, the metrics, and the research questions.}
\end{figure*}
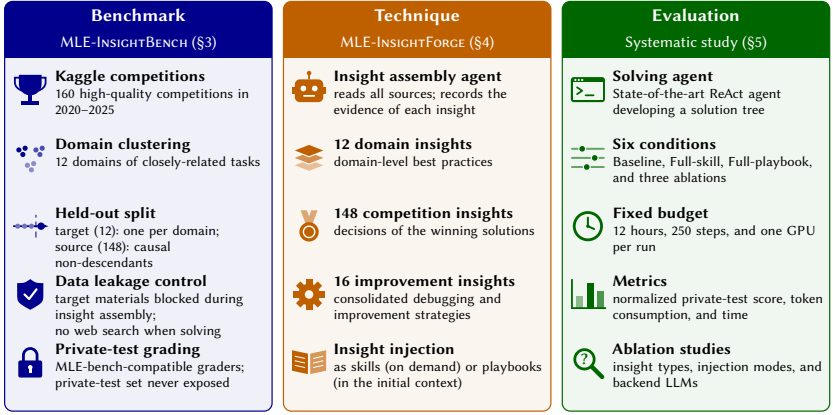

To bridge these gaps, we systematically study how insights assembled from closely-related \MLE tasks affect \AgenticMLESystems, and how their effect depends on the type of insights and the way they are injected. Figure~\ref{fig:overview} illustrates the workflow of our study.
We first build \Benchmark from Kaggle competitions with human-written solutions, i.e., gold-medal teams' notebooks and writeups, and group them into domains of closely-related competitions.
One competition of each domain is held out as a \TargetTask, and only the remaining \UseMacro{N-CorpusCompetitions} competitions, all \NonDescendants of the \TargetTasks (i.e., none can have been influenced by a \TargetTask, as a later edition of the same series or a competition reusing its solutions could), serve as \SourceCompetitions, which mirrors practice and rules out insights derived from a \TargetTask's own winning solutions.

\Tool then reads the descriptions, datasets, solution notebooks, and writeups of the \SourceCompetitions and assembles three types of insights: 
\CompetitionInsights on the decisions behind each competition's winning solutions, \DomainInsights that reconcile them into domain-level best practices, and \ImprovementInsights that connect problems encountered during development to their resolutions.
Each insight consists of \TriggerAction items that a system can skip when they do not apply, and is backed by records of its source passages.
The insights can be injected as \Skills, whose full text is read on demand~\cite{Anthropic25AgentSkills}, or as a \Playbook in the initial context~\cite{ZhangETAL26ACE}.

On the \UseMacro{N-EvalCompetitions} \TargetTasks, the assembled insights improve the normalized score of an \AgenticMLESystem by \UseMacro{N-SkillGain} over the baseline without insights, improving every task (significantly on \UseMacro{N-SkillTasksSignificant}) when injected as \Skills; on an energy forecasting task, every run with insights beats the best run without them, although none of its peer competitions involves energy.
Further controlled ablation studies show that the three types of insights are complementary: each type alone significantly improves fewer tasks (\UseMacro{N-DomainPlaybookTasksSignificant}, \UseMacro{N-CompetitionPlaybookTasksSignificant}, and \UseMacro{N-ImprovementPlaybookTasksSignificant}) than their combination (\UseMacro{N-PlaybookTasksSignificant}).
The improvements from insights generalize across backend LLMs, including a stronger proprietary LLM (\GPTModel).
However, the better injection mode varies across tasks, with \Skills being more cost-effective overall, because they reach a similar score with fewer input tokens (\UseMacro{N-SkillInputTokensChange} versus \UseMacro{N-PlaybookInputTokensChange} more than the baseline).

\noindent
The main contributions of this work are threefold:

\begin{itemize}[topsep=2pt,itemsep=0.3ex,partopsep=0ex,parsep=0ex,leftmargin=*]
\item
\MyPara{Benchmark}
We collect \Benchmark, a benchmark of \UseMacro{N-BenchmarkCompetitions} Kaggle competitions clustered into \UseMacro{N-Domains} domains, that provides a controlled environment for studying insights: its held-out split makes the sources of insights explicit and keeps \TargetTask information out of them.
\item
\MyPara{Technique}
We design and implement \Tool, an agent that assembles evidence-based \DomainInsights, \CompetitionInsights, and \ImprovementInsights from human-written solutions and writeups, and injects them into \AgenticMLESystems as \Skills or \Playbooks.
\item
\MyPara{Experiments}
We systematically evaluate the impact of assembled insights on an \AgenticMLESystem, including their overall effectiveness, the contributions of the three types of insights, the effectiveness and cost of the two injection modes, and the generalizability across backend LLMs.
\end{itemize}

% \noindent
% Our artifact, including \Benchmark, \Tool, and the assembled insights, is available on FigShare~\cite{Artifact}.

\section{Background}
\label{sec:background}

\subsection{\AgenticMLECap Systems}
\label{sec:background:systems}

\begin{figure}[t]
\centering
\resizebox{0.9\textwidth}{!}{\input{figs/example-task}}
\vspace{-4pt}
\caption{An example Kaggle competition and a simplified ML solution developed by \Baseline.}\label{fig:example-task}
\vspace{-6pt}
\Description{Two side-by-side boxes. (a) The task Jigsaw -- Agile Community Rules Classification, styled as its Kaggle competition page with the address kaggle.com/competitions/jigsaw-agile-community-rules, the subtitle "Using AI models to help moderators uphold community-specific norms", and the sections Description, Dataset, Metric, and Submission: the description asks for a binary classifier that predicts whether a Reddit comment broke a specific rule, with unseen rules in the test data; the dataset consists of train.csv and test.csv, illustrated by one row with the comment body, the rule, the subreddit, positive and negative example comments, and the binary target rule\_violation; the metric is the column-averaged AUC; the submission file contains a probability for each row. (b) A Python solution in four stages: data preparation, which encodes the comment, the rule, and the example comments with a sentence-embedding model and computes their similarities as features; validation, which trains a logistic regression with each rule held out in turn and reports the mean AUC; training on all data; and test, which predicts the test data and writes submission.csv.}
\end{figure}

An \MLE task asks for an executable ML solution rather than only a fitted model: it involves understanding the data and the metric, designing a validation protocol, implementing preprocessing and training, diagnosing failures, and producing an artifact that satisfies a submission contract.
These decisions are coupled: for example, the validation split must reflect the deployment distribution, while model and feature choices must respect runtime and memory limits.
Competitions on Kaggle~\cite{Kaggle} expose \MLE tasks through a common interface consisting of a task description, datasets, an evaluation metric, a compute environment, and a submission format.
Figure~\ref{fig:example-task} shows an example Kaggle competition, Jigsaw -- Agile Community Rules Classification~\cite{KaggleJigsawRules}, which asks whether a Reddit comment violates a given rule, including rules unseen in training, together with a solution developed by an \AgenticMLESystem.
The solution prepares features from the training data, validates the model by holding out one rule at a time to mimic the unseen rules, retrains on all training data, and writes the predictions for the test data in the submission format.

\MLEBench~\cite{ChanETAL25MLEBench} turns 75 Kaggle competitions into an offline benchmark: each competition's training data is split into new training and private-test sets, and a submission is graded against the private-test labels and compared with the human leaderboard.
MLRC-Bench~\cite{ZhangETAL25MLRCBench} instead targets research competitions held at ML conferences (e.g., NeurIPS and KDD Cup), but contains only seven tasks.
Neither benchmark specifies a temporal or causal ordering among the tasks, i.e., there is no standard of which other tasks may serve as sources of insights for a given task.

An \AgenticMLESystem combines an LLM with tools for inspecting data, editing code, executing experiments, and observing validation feedback, typically following the ReAct pattern of interleaved reasoning and tool use~\cite{YaoETAL23ReAct}.
Beyond a single linear trajectory, most recent systems search over candidate solutions.
AIDE~\cite{JiangETAL25AIDE} maintains a tree of solutions that are drafted, debugged, and improved; ML-Master~\cite{LiuETAL25MLMaster} explores such trees with Monte Carlo tree search, and AIRA-dojo~\cite{ToledoETAL25AIRADojo} compares search policies and operators over them; and FM Agent and LoongFlow~\cite{LiETAL25FMAgent,WanETAL25LoongFlow} evolve populations of solutions.
Within a task, these systems accumulate the code, scores, and reflections produced by earlier candidates to guide later ones.

\subsection{Insights in \AgenticMLECap Systems}
\label{sec:background:insights}

We use \emph{\textbf{insight}} to denote a reusable, natural-language statement about how to solve \MLE tasks, derived from tasks other than the one being solved.
Recent systems obtain insights in two ways:

\begin{itemize}[leftmargin=*]
\item \textbf{External knowledge:}
Insights are curated from external sources, such as solutions of similar MLE tasks on open-source repositories and competition platforms, API documentations, and forum discussions.
Some techniques retrieve them on the fly, typically via web search (e.g., MLE-STAR~\cite{NamETAL25MLESTAR} and MARS~\cite{ChenETAL26MARS});
others build a knowledge base offline which is used for retrieval-augmented generation (RAG) at solving time (e.g., DS-Agent~\cite{GuoETAL24DSAgent}, MLEvolve~\cite{DuETAL26MLEvolve}, AutoMLGen~\cite{DuETAL25AutoMLGen}, KAPSO~\cite{NadafianETAL26KAPSO}, AIBuildAI-2~\cite{ZhangETAL26AIBuildAI2}, and FM Agent~\cite{LiETAL25FMAgent}).
% human-written materials of peer tasks, either at solving time or offline.
% Solving time examples include: MLE-STAR which seeds its initial solutions with models and example code retrieved by web search~\cite{NamETAL25MLESTAR}, and MARS which uses web search to identify candidate architectures and to debug~\cite{ChenETAL26MARS}.
% Offline examples include DS-Agent, MLEvolve, AutoMLGen, KAPSO, AIBuildAI-2, and FM Agent, which build knowledge bases from open-source repositories, documentation, and competition platforms and forums, and retrieve them during solving~\cite{DuETAL26MLEvolve,DuETAL25AutoMLGen,NadafianETAL26KAPSO,ZhangETAL26AIBuildAI2,LiETAL25FMAgent}.
%DS-Agent~\cite{GuoETAL24DSAgent} earlier retrieved manually curated Kaggle cases for case-based reasoning.
\item \textbf{Self-improvement across tasks:}
Insights are distilled from the system's own runs on previous tasks, usually generated by the agent itself or a separate surrogate agent.
Examples include the accumulated memory of Agent K~\cite{GrosnitETAL24AgentK}, the ``prior wisdom'' cache of ML-Master 2.0~\cite{ZhuETAL26MLMaster2}, and the takeaways of completed runs of AIBuildAI-2~\cite{ZhangETAL26AIBuildAI2} (which augments its external knowledge base).
%  general-purpose agents similarly evolve their context from execution feedback~\cite{ZhangETAL26ACE}.
\end{itemize}

% The insights we study belong to the first category, but are assembled offline from the human-written solutions of a fixed set of \SourceCompetitions that excludes the \TargetTasks (\S\ref{sec:benchmark}).}

Regardless of their source, insights reach the agent as text in its context, and systems differ in when this happens.
In the \emph{skill} mode~\cite{Anthropic25AgentSkills}, only compact metadata of each insight (a name and a description) is placed in the context, and the agent loads the full text on demand, a design known as progressive disclosure.
In the \emph{playbook} mode~\cite{ZhangETAL26ACE}, the insights are placed directly in the initial context as a playbook; unlike a human-written prompt, a playbook is generated and maintained by LLM agents.
Skills save context but depend on the agent retrieving the right insight, whereas playbooks make every insight available but consume context even for irrelevant insights.

\subsection{Data Leakage in Evaluating \AgenticMLECap Systems}
\label{sec:background:leakage}

An evaluation of an \AgenticMLESystem suffers from data leakage when the system has access, during solving, to information about the \TargetTask that would not be available when solving a genuinely new task, most notably the task's own winning solutions.
This differs from train--test leakage in a generated solution, which some systems check~\cite{NamETAL25MLESTAR,DuETAL26MLEvolve}, and from contamination of the backend LLM's pre-training data, which several papers acknowledge as a limitation~\cite{NamETAL25MLESTAR,ToledoETAL25AIRADojo}.

Insights are a direct channel for such leakage, because their sources are usually the same public competition materials from which \TargetTasks are drawn.
Web search at solving time is not reported to be restricted to materials of other tasks~\cite{NamETAL25MLESTAR,ChenETAL26MARS}, and knowledge bases built from competition platforms typically do not report whether the \TargetTasks were excluded~\cite{DuETAL26MLEvolve,DuETAL25AutoMLGen}.
Some work did exclude the \TargetTasks by competition name~\cite{ZhuETAL26MLMaster2,YangETAL25RDAgent}, which does not cover later editions of the same competition series.
For example, \Benchmark contains two AI Mathematical Olympiad: Progress Prize 1~\cite{KaggleAIMO1} and Progress Prize 2~\cite{KaggleAIMO2}; if the first were a \TargetTask, excluding it by name would still admit the second, whose description states that it builds upon the first and one of whose gold-medal notebooks reuses code from the winning solution of the first.

Cross-task memory raises a related concern about the order of tasks.
Accumulating experience over a sequence of tasks, i.e., online learning, is not leakage by itself, as it mirrors how practitioners gain experience over time.
However, when the memory keeps accumulating across the \TargetTasks themselves, a task may be solved after another task that it influenced. %, such as a later edition of its series, whose dataset may include the earlier edition's test data and whose solutions may build on the earlier edition's winning solutions.
Avoiding leakage thus requires a chronological or causal ordering of the tasks.
% the experience available when solving a task to come only from tasks that cannot have been influenced by it. %, which holds, for instance, when the tasks are solved in chronological order.
These practices have been raised as a fairness concern on the \MLEBench leaderboard~\cite{MLEBenchIssue138}, which is one of the reasons that the leaderboard has paused new submissions.
Contamination-free benchmarks in code generation address a similar problem by evaluating only on problems released after the model's training cutoff~\cite{JainETAL25LiveCodeBench}; \Benchmark (\S\ref{sec:benchmark}) applies a similar principle to insight sources by using only competitions that cannot have been influenced by the \TargetTasks.

\section{\Benchmark: A Controlled-Environment Benchmark}
\label{sec:benchmark}

In \Benchmark, the \SourceCompetitions are fixed and disjoint from the \TargetTasks: competitions are grouped into domains so that every \TargetTask has closely-related peer tasks, and the split is fixed before any insight is assembled.

\begin{wrapfigure}{r}{0.36\textwidth}
\centering
\includegraphics[width=\linewidth]{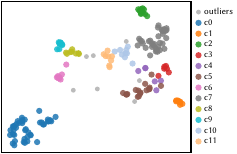}
\caption{UMAP embeddings of the competition descriptions (\UseMacro{N-DescriptionClusters} clusters).}\label{fig:competition-clusters}
\Description{A scatter plot of UMAP embeddings of the competition descriptions, clustered into twelve clusters plus noise.}
\end{wrapfigure}
\MyPara{Competition collection}
We collect Kaggle competitions using \MetaKaggle~\cite{KaggleMetaKaggle} and Meta Kaggle Code~\cite{KaggleMetaKaggleCode}.
We retain competitions that are launched in 2020 or later, use Python, have downloadable datasets, and have high-performing self-contained solutions, and exclude playground competitions; a solution counts as high-performing and self-contained if it received a gold medal and uses no datasets other than the competition's own.
This step yields \UseMacro{N-MinedCompetitions} competitions, from which we manually remove \UseMacro{N-NonMLCompetitions} that do not involve training ML models (simulation competitions and optimization puzzles, such as the Santa series~\cite{santa-2025}), leaving \UseMacro{N-ClusteredCompetitions} candidate competitions. % , and competitions without a dataset
For each competition, we collect the task description, the public writeups of the teams ranked within the gold-medal zone of the leaderboard, and the high-performing self-contained solution notebooks; we refer to the latter two as the competition's human-written solutions.

\begin{table}[t]
\begin{small}
\begin{center}
\caption{\TargetTasksCap of \Benchmark.}\label{tab:benchmark}
%% Auto-generated by automation/generate_benchmark_table.py from LLM-Insight-Graph@fddb0b72.
\begin{tabular}{@{}ll@{\hspace{0.8em}}ll@{}}
\toprule
\UseMacro{TH-Domain} & \UseMacro{TH-TargetTask} & \UseMacro{TH-Metric} & \UseMacro{TH-Type} \\
\midrule
Agriculture Recognition & CSIRO Biomass~\cite{csiro-biomass} & $R^2$$\uparrow$ & Simple \\
Bioinformatics & CAFA 5 Protein Function~\cite{cafa-5-protein-function-prediction} & F$_\text{max}$$\uparrow$ & Simple \\
Geography Retrieval and Localization & Image Matching 2025~\cite{image-matching-challenge-2025} & mAA$\uparrow$ & Simple \\
Medical Image Recognition & RSNA Aneurysm Detection~\cite{rsna-intracranial-aneurysm-detection} & AUC$\uparrow$ & Code \\
NLP & Jigsaw Community Rules~\cite{jigsaw-agile-community-rules} & AUC$\uparrow$ & Simple \\
Physiological Signal Processing & PhysioNet ECG Digitization~\cite{physionet-ecg-image-digitization} & SNR$\uparrow$ & Simple \\
Quantitative Finance & Mitsui Commodity Prediction~\cite{mitsui-commodity-prediction-challenge} & Sharpe$\uparrow$ & Code \\
Scientific Signal Processing & Ariel Data Challenge 2025~\cite{ariel-data-challenge-2025} & GLL$\uparrow$ & Simple \\
Sport Tournament Prediction & March Mania 2025~\cite{march-machine-learning-mania-2025} & Brier$\downarrow$ & Simple \\
Time Series Prediction & Energy Behavior of Prosumers~\cite{KaggleEnefit} & MAE$\downarrow$ & Code \\
Video Recognition & NFL Player Contact~\cite{nfl-player-contact-detection} & MCC$\uparrow$ & Simple \\
Other & POG Series S01E01~\cite{kaggle-pog-series-s01e01} & MAE$\downarrow$ & Simple \\
\bottomrule
\end{tabular}

\end{center}
\end{small}
\end{table}

\MyPara{Domain clustering}
We group the candidate competitions by clustering their natural-language descriptions (overview and dataset description), and then manually assign a human-readable label to each cluster.
We embed each description with \EmbeddingModel~\cite{Microsoft26Harrier}, a text embedding model, by instructing it to cluster competitions by ML task formulation and application domain.
We then reduce the embeddings to 11 dimensions with UMAP~\cite{McInnesETAL18UMAP} (cosine distance, 20 neighbors) and cluster them with HDBSCAN~\cite{CampelloETAL13HDBSCAN} (minimum cluster size 5, minimum samples 3), which determines the number of clusters from the data and leaves outliers unassigned.
This yields \UseMacro{N-DescriptionClusters} clusters and \UseMacro{N-ClusterOutliers} outliers (unable to be clustered), as shown in Figure~\ref{fig:competition-clusters}.
We label each cluster by inspecting its member competitions and its most distinctive terms under class-based TF-IDF~\cite{Grootendorst22BERTopic}, and manually move the outlier WiDS Datathon 2022~\cite{widsdatathon2022} into the cluster that contains its other editions.
Domains serve two purposes: they define the scope of \DomainInsights, and they ensure that every \TargetTask has closely-related peer tasks from which insights can be assembled.
We exclude the Audio Recognition cluster, where none of its \UseMacro{N-AudioCompetitions} competitions satisfies the eligibility rule for \TargetTasks described below, as well as \UseMacro{N-NoWriteupCompetitions} other competitions without any writeup from the gold-medal zone, and Jane Street Market Prediction~\cite{jane-street-market-prediction}, whose dataset was available at our initial collection but is no longer publicly accessible.
The resulting \Benchmark consists of \UseMacro{N-BenchmarkCompetitions} competitions in \UseMacro{N-Domains} domains, including a residual domain labeled \emph{Other} for the remaining outliers.
% The cluster ids 2 and 3 carry swapped labels in the main repo (Assets/, Insights/, mlebench configs);
% the labels are only names, and automation/generate_benchmark_table.py maps them to the correct
% domain names. Confirmed with Bihui; do not discuss the swap in the paper.

\MyPara{Held-out split}
From each domain, we hold out one eligible competition as a \TargetTask and use the remaining \UseMacro{N-CorpusCompetitions} competitions as \SourceCompetitions.
A competition is eligible if the mean runtime of its collected solutions is at most the 75th percentile of the runtimes of all collected solution notebooks (\UseMacro{N-RuntimeThresholdMin} minutes), which keeps the evaluation within a fixed compute budget.
We also manually mark some competitions as ineligible where the executable environment for the tasks are hard to replicate (e.g., using a huge dataset): AMP Parkinson's Disease Progression Prediction~\cite{amp-parkinsons-disease-progression-prediction}, Drawing with LLMs~\cite{drawing-with-llms}, Bengali.AI Speech Recognition~\cite{bengaliai-speech}, Cornell Birdcall Identification~\cite{birdsong-recognition}, and the BirdCLEF series (e.g., BirdCLEF+ 2025~\cite{birdclef-2025}).
% Decided not to state why each of these competitions is ineligible; the list suffices.
Among the eligible competitions of the most recent year in a domain, we select the one with the shortest solution runtime.

The resulting \NonDescendantSplit ensures that every \SourceCompetition is a \NonDescendant of its \TargetTask, i.e., a competition that cannot have been influenced by the \TargetTask as a later edition of the same series or reusing its solutions.
This mirrors how insights would be used in practice: an \AgenticMLESystem facing a new task can draw on prior experience, but not on experience derived from the new task itself.
While a chronological split (i.e., choosing the latest competition of each domain) is more straightforward, it would reduce the size of benchmark because the latest competitions sometimes are not eligible as a \TargetTask, causing dropping of valuable \SourceCompetitions that could otherwise be used with our more careful split.
% A \TargetTask is thus not necessarily the latest competition of its domain: because ineligible competitions remain \SourceCompetitions, \UseMacro{N-SameYearSources} sources are launched in the same year as their \TargetTask and \UseMacro{N-LaterSources} (in \UseMacro{N-LaterSourceDomains} domains) in a later year, but none of them belongs to the series of its \TargetTask or refers to it in its description, writeups, or solutions.
Table~\ref{tab:benchmark} lists the \UseMacro{N-EvalCompetitions} \TargetTasks, of which \UseMacro{N-EvalSimple} are standard competitions with a submission file and \UseMacro{N-EvalCode} are code competitions whose predictions are produced through a competition-specific evaluation API.

\MyPara{Leakage control}
\Benchmark controls the sources of insights at three points.
First, the materials of the \TargetTasks (datasets, descriptions, writeups, and solutions) are excluded from insight assembly, so no assembled insight can be derived from a \TargetTask's own winning solutions.
Second, because the split only requires the \SourceCompetitions to be \NonDescendants of the \TargetTask, earlier editions of a recurring competition series (e.g., March Machine Learning Mania) remain legitimate \SourceCompetitions, as they would be for a practitioner. %; to measure how much such closely-related editions contribute, \Benchmark also supports a stricter leave-family-out split that removes every edition of the \TargetTask's series.
Third, the \AgenticMLESystem under evaluation is not given web-search tools and receives only the public data of the \TargetTask; the private-test inputs are substituted only when the final solution is graded (same as how Kaggle grades submissions), and the private-test labels are never exposed.
% The web-access audit of the run trajectories (N-Audit* macros from
% automation/generate_trajectory_stats.py) found a few runs that fetched participants' code of the
% target task from GitHub; their final solutions adopt only metric code. Decided not to report
% the audit in the paper (including in RQ4); do not add it back.
% Grading follows the \MLEBench protocol~\cite{ChanETAL25MLEBench}: we port each \TargetTask into an \MLEBench-compatible preparation and grading script and score the solution against the private-test labels.
% \Fix{XXX: Freeze and publish the leave-target-out snapshots (with hashes) used to assemble the insights for each \TargetTask.}

\section{Methodology}
\label{sec:technique}

\subsection{\Tool: Insight Assembly Agent}
\label{sec:technique:tool}

\Tool explores the descriptions, datasets, top human-written solutions, and writeups of the \SourceCompetitions and assembles three types of insights from them.
Separating \emph{what} insights are assembled from \emph{how} they are injected (\S\ref{sec:technique:usage}) lets us vary the two independently.

\subsubsection{Agent Architecture}
\label{sec:technique:agent}

\begin{figure*}[t]
\centering
\resizebox{\textwidth}{!}{\begin{tikzpicture}[
  x=1cm, y=1cm, font=\sffamily\scriptsize,
  box/.style={draw=black!55, fill=white, rounded corners=2pt, align=center, inner sep=3pt,
    execute at begin node={\hyphenpenalty=10000\exhyphenpenalty=10000}},
  lbox/.style={box, align=left},
  doc/.style={box, shape=tape, rounded corners=0pt, tape bend top=none, tape bend height=4pt,
    text width=1.6cm, minimum height=0.8cm},
  stage/.style={box, draw=MainColor, fill=MainFill, text width=1.45cm, minimum height=0.8cm},
  stack/.style={box, double copy shadow={shadow xshift=2.5pt, shadow yshift=-2.5pt},
    text width=2.5cm, minimum height=0.8cm},
  ins/.style={stack, draw=#1Color, fill=#1Fill, text=#1Color!70!black},
  msg/.style={-{Latex[length=1.6mm]}, line width=0.6pt},
  ret/.style={-{Latex[length=1.6mm]}, densely dashed, line width=0.6pt},
  lbl/.style={inner sep=1.5pt, align=center, font=\sffamily\tiny},
  ptitle/.style={font=\sffamily\small\bfseries, anchor=west, inner sep=0pt},
]
%% (a) agent
\node[ptitle] at (0,0.35) {(a) Assembly agent};

\node[doc, text width=1.3cm] (task) at (0.8,-2.75) {\textbf{Assembly task}};

\fill[MainFill, rounded corners=4pt] (1.85,-1.4) rectangle (8.75,-4.7);
\node[anchor=north west, font=\sffamily\scriptsize\itshape, MainColor] at (1.9,-1.43) {ReAct loop};

\pic[scale=0.8] at (3.1,-2.0) {ov-robot};
\node[box, draw=MainColor, text width=1.0cm] (llm) at (3.1,-2.75) {\textbf{LLM}};
\node[box, text width=0.95cm] (sum) at (2.65,-4.05) {context summary};

\node[lbox, text width=2.4cm] (tools) at (7.1,-2.45) {\textbf{Tools}\\[1pt]
  files, notebooks\\ datasets, sandbox\\ web, PDFs, repos\\ images, videos};
\node[lbox, text width=2.4cm] (gate) at (7.1,-4.05) {\textbf{Validator}\\[1pt] full reads, evidence};

\draw[msg] (task.east) -- (llm.west);
\draw[msg] ([yshift=0.15cm]llm.east) -- ([yshift=0.15cm]llm.east -| tools.west) node[lbl, midway, above] {tool calls};
\draw[ret] ([yshift=-0.15cm]llm.east -| tools.west) -- ([yshift=-0.15cm]llm.east) node[lbl, midway, below] {observations};
\draw[ret, {Latex[length=1.6mm]}-{Latex[length=1.6mm]}] ([xshift=-0.45cm]llm.south) -- (sum.north);
\draw[msg] ([xshift=0.2cm]llm.south) |- ([yshift=0.12cm]gate.west) node[lbl, pos=0.75, above] {submit};
\draw[ret] ([yshift=-0.12cm]gate.west) -| ([xshift=0.45cm]llm.south) node[lbl, pos=0.25, below] {reject};

\node[box, shape=cylinder, shape border rotate=90, aspect=0.2, rounded corners=0pt,
  minimum width=2.0cm, minimum height=0.95cm] (ws) at (5.2,-0.45) {\textbf{Workspace}\\read-only};
\node[box, shape=cloud, cloud puffs=11, aspect=2.4, rounded corners=0pt, inner sep=0pt,
  minimum width=2.1cm, minimum height=0.95cm] (web) at (7.9,-0.45) {\textbf{Web}\\filtered};
\draw[msg] ([xshift=-0.9cm]tools.north) -- ([xshift=-0.9cm]tools.north |- ws.south);
\draw[msg] ([xshift=0.8cm]tools.north) -- ([xshift=0.8cm]tools.north |- web.south);

\node[stack, draw=MainColor, fill=MainFill, text width=2.3cm] (out) at (7.05,-5.45) {insights + evidence};
\draw[msg] (gate.south) -- (out.north);

\draw[black!30, densely dashed, line width=0.5pt] (9.15,0.45) -- (9.15,-6.0);

%% (b) workflow
\node[ptitle] at (9.5,0.35) {(b) Assembly workflow};

\node[doc] (in1) at (10.55,-0.8) {source competition};
\node[stage] (s1) at (12.8,-0.8) {\textbf{1.} Assemble};
\node[ins=Competition] (o1) at (15.45,-0.8) {\textbf{\UseMacro{N-CompetitionInsights}} \CompetitionInsights};

\node[doc] (in2) at (10.55,-2.3) {domain descriptions};
\node[stage] (s2) at (12.8,-2.3) {\textbf{2.} Synthesize};
\node[ins=Domain] (o2) at (15.45,-2.3) {\textbf{\UseMacro{N-Domains}} \DomainInsights};

\node[doc] (in3) at (10.55,-3.8) {problem--fix writeups};
\node[stage] (s3) at (12.8,-3.8) {\textbf{3.} Assemble};
\node[ins=Improvement, densely dashed] (o3) at (15.45,-3.8) {\textbf{\UseMacro{N-ImprovementInsights}} \RawImprovementInsights};

\node[stage] (s4) at (12.8,-5.3) {\textbf{4.} Cluster, synthesize};
\node[ins=Improvement] (o4) at (15.45,-5.3) {\textbf{\UseMacro{N-SynthesizedImprovements}} \ImprovementInsights};

\foreach \i in {1,2,3} {
  \draw[msg] (in\i.east) -- (s\i.west);
  \draw[msg] (s\i.east) -- (o\i.west);
}
\draw[msg] (s4.east) -- (o4.west);
\draw[msg] ([yshift=-2.5pt]o1.south) -- ++(0,-0.25) -| (s2.north) node[lbl, pos=0.25, below] {per domain};
\draw[msg] ([yshift=-2.5pt]o3.south) -- ++(0,-0.25) -| (s4.north) node[lbl, pos=0.25, below] {clustered};
\end{tikzpicture}}
\vspace{-4pt}
\caption{Architecture of \Tool: (a) the assembly agent and (b) the four assembly stages.}\label{fig:assembler}
\vspace{-6pt}
\Description{Two panels. Panel (a) shows an assembly task, consisting of instructions and required sources, entering a ReAct loop in which an LLM calls tools (reading files and notebooks, inspecting datasets, running analyses in a sandbox, reading web pages, PDFs, and repositories, and inspecting images and videos) and receives observations; the tools access a read-only workspace with the \SourceCompetitions and the blacklist-filtered web; the LLM's context is summarized when needed; the LLM submits a report to a validation step that checks that all required sources were read in full and that each insight has an evidence record, returning rejections to the LLM and otherwise producing the insight text and evidence records. Panel (b) shows four stages: competition insight assembly from each \SourceCompetition, domain insight synthesis from the competition insights and descriptions of a domain, improvement insight assembly from writeups reporting a problem and its resolution, and clustering and synthesis of the raw improvement insights into improvement insights.}
\end{figure*}
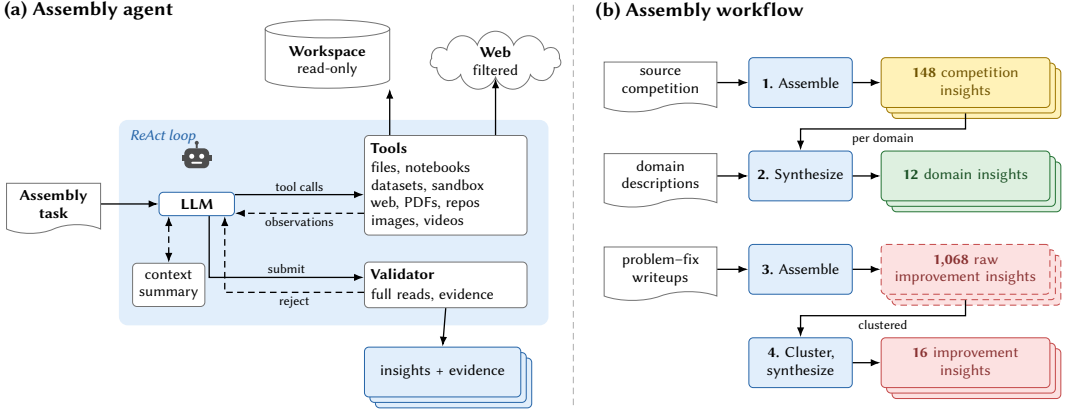

\Tool (Figure~\ref{fig:assembler}(a)) is a ReAct-style agent~\cite{YaoETAL23ReAct}, using \AssemblyModel as its backend LLM during our study.
Each run of the agent performs one assembly task, specified by the instructions of its stage and the list of source files it must read.
The agent's tools read files and notebooks, inspect datasets, run targeted analyses in a sandboxed container (e.g., checking a preprocessing step or a validation split claimed in a writeup, without rerunning the solution end-to-end), and read the materials linked from writeups, such as web pages, PDFs, code repositories, images, and videos.
% When the conversation history approaches the context window of the LLM, older messages are replaced by a summary.

\MyPara{Leakage prevention}
The source materials are mounted read-only in the agent's workspace, which contains only the \SourceCompetitions and not the \TargetTasks.
Because some URLs in the source materials may also link to the \TargetTasks, web access goes through a blacklist of \UseMacro{N-BlacklistEntries} URL patterns, which we extract from the descriptions, writeups, and solution notebooks of the \TargetTasks; links to Kaggle or GitHub the competition's or repository's all sub-pages, while general documentation and tutorial sites (e.g., library documentation and Wikipedia) are manually exempted.

\MyPara{Evidence-based submission}
The agent submits each insight through a validated submission interface, which rejects the submission until the agent has read every required source in full, otherwise returns the incomplete insight to the agent for repair.
To minimize hallucination, the interface also requires an evidence record for every insight, listing the original source passages it is based on; this ensures that every insight can be traced back to its original source.

\MyPara{Assembly stages}
Figure~\ref{fig:assembler}(b) shows the four stages in which \Tool runs.
It first assembles one \CompetitionInsight per \SourceCompetition and then synthesizes one \DomainInsight per domain from the \CompetitionInsights of that domain.
Independently of these two types, it assembles raw \ImprovementInsights from the writeups and consolidates them by clustering.

\subsubsection{Insight Types}
\label{sec:technique:types}

\Tool assembles \UseMacro{N-CompetitionInsights} \CompetitionInsights, \UseMacro{N-Domains} \DomainInsights, and \UseMacro{N-SynthesizedImprovements} \ImprovementInsights, at a total API cost of \$\UseMacro{N-AssemblyCost} USD at the list prices of \AssemblyModel (\$\UseMacro{N-AssemblyCostCompetition}, \$\UseMacro{N-AssemblyCostDomain}, and \$\UseMacro{N-AssemblyCostImprovement} USD for the three types, respectively).

\MyPara{\CompetitionInsightsCap}
\Tool first reads the description to identify the objective, data, metric, and constraints, and then reads every collected writeup in full.
It consults the solution notebooks to disambiguate implementation details and to check that a technique claimed in a writeup is actually present, and inspects the datasets to understand their characteristics and to check claims about them.
The resulting insight is organized around the decisions that distinguished the winning solutions and states the conditions under which each decision was made, rather than reproducing a particular solution, since the insight is used for other tasks, to which a complete solution tailored to the \SourceCompetition's data and metric does not transfer.

\MyPara{\DomainInsightsCap}
A \TargetTask has no insight of its own, but it shares task properties with the other competitions in its domain.
\Tool reads the descriptions and \CompetitionInsights of all \SourceCompetitions in the domain and reconciles recommendations that recur, complement each other, or conflict.
It replaces competition-specific prescriptions with conditional guidance; for example, it associates a validation strategy with grouped subjects or temporal deployment rather than as universally applicable.
A \DomainInsight drops recommendations specific to a single competition, keeping only those supported by several competitions or a task archetype.

\begin{wrapfigure}{r}{0.36\textwidth}
\centering
\includegraphics[width=\linewidth]{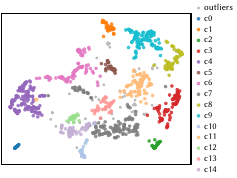}
\caption{UMAP embeddings of the \RawImprovementInsights (\UseMacro{N-ImprovementClusters} clusters).}\label{fig:improvement-clusters}
\Description{A scatter plot of UMAP embeddings of the raw improvement insights, clustered into groups plus noise.}
\end{wrapfigure}
\MyPara{\ImprovementInsightsCap}
While the other two types describe the decisions behind the final solutions, \ImprovementInsights capture how their authors diagnosed and fixed problems along the way, by connecting an observed problem (e.g., a validation split that does not match the deployment setting, or an inference pipeline that exceeds the runtime budget) to the resolution that fixed it.
\Tool first filters writeups, with an LLM, for those that report both a concrete problem and an actionable resolution.
For each retained writeup, it then assembles standalone insights, keeping only those whose outcome is directly reported in the writeup; this step yields \UseMacro{N-ImprovementInsights} \RawImprovementInsights.
Because many improvement strategies recur across competitions, exposing all of them individually would make them costly to discover and highly redundant.
We therefore cluster the \RawImprovementInsights by their descriptions, using the same approach as Domain Clustering (\S\ref{sec:benchmark}), and let \Tool synthesize each cluster into one consolidated insight under the same evidence constraints.
Specifically, we embed the text of each raw insight with \EmbeddingModel, using an instruction to identify the prediction task type, data modality, and application domain that the insight targets, and cluster the embeddings with UMAP (15 dimensions, 20 neighbors) and HDBSCAN (minimum cluster size 20, minimum samples 5); we select these hyperparameters by a grid search for clusters that are stable across random seeds, align with the data modality and solution stage of the insights, and cover most insights.
This yields \UseMacro{N-ImprovementClusters} clusters, and the outliers (\UseMacro{N-ImprovementOutlierShare} of the raw insights) form the residual cluster, as shown in Figure~\ref{fig:improvement-clusters}.
\Tool then synthesizes each cluster into one consolidated \ImprovementInsight, including one for the residual cluster, resulting in \UseMacro{N-SynthesizedImprovements} \ImprovementInsights.
A manifest records the member insights and evidence of each cluster, so that every \ImprovementInsight remains traceable to its raw ones.

\subsubsection{Insight Representation}
\label{sec:technique:representation}

A \CompetitionInsight or \DomainInsight consists of \InsightItems following the \TriggerAction pattern: a trigger (a task or data condition that can be checked before acting), one recommended action, and a rationale for its expected effect.
Triggers are phrased in terms of general task and data properties, so the \AgenticMLESystem can skip the items that do not apply to its task; conflicting findings are kept as items with mutually exclusive triggers.
The items are grouped by five \Aspects of an ML solution (data modality, data, model, training, and other) and ordered within each aspect by the cost of following them.
An \ImprovementInsight is a single item whose trigger is the observed problem and whose action is the resolution, elaborated with its limitations, cost, and how to check whether it worked.
Each insight has a name and a description; its evidence records are stored separately and enter the context only if explicitly read.

\subsection{Usage of Insights}
\label{sec:technique:usage}

\subsubsection{Base Agentic MLE System}
\label{sec:technique:system}

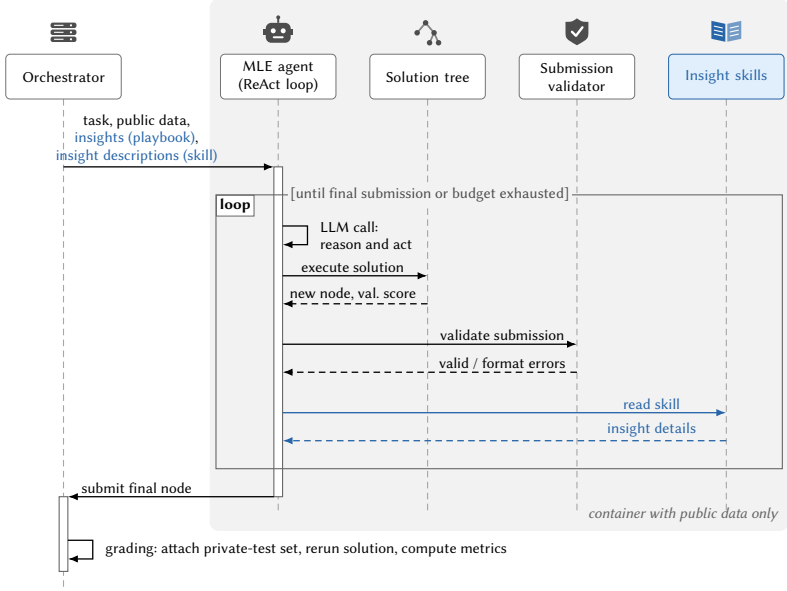
\begin{figure*}[t]
\centering
\resizebox{0.75\textwidth}{!}{\begin{tikzpicture}[
  x=1cm, y=1cm, font=\sffamily\scriptsize,
  head/.style={draw=black!55, fill=white, rounded corners=2pt, align=center, text width=1.75cm,
    minimum height=0.72cm, inner sep=1.5pt},
  life/.style={black!35, densely dashed, line width=0.5pt},
  act/.style={draw=black!60, fill=white, line width=0.5pt},
  msg/.style={-{Latex[length=1.6mm]}, line width=0.6pt},
  ret/.style={-{Latex[length=1.6mm]}, densely dashed, line width=0.6pt},
  lbl/.style={inner sep=1.5pt, align=center, anchor=south},
  ins/.style={MainColor},
]
\def\xO{0}
\def\xA{3.45}
\def\xT{5.85}
\def\xV{8.25}
\def\xS{10.65}
\def\yCtr{-7.0}
\def\yEnd{-8.25}
\newcommand{\SelfCall}[2]{%
  \draw[msg] ({#1+0.07},#2) -- ++(0.4,0) -- ++(0,-0.3) -- ++(-0.4,0);
}

% container
\fill[black!5, rounded corners=4pt] ({\xA-1.1},1.25) rectangle ({\xS+1.0},-7.3);
\node[anchor=south east, font=\sffamily\scriptsize\itshape, black!60] at ({\xS+0.95},-7.28) {container with public data only};

% lifelines
\draw[life] (\xO,-0.36) -- (\xO,\yEnd);
\foreach \x in {\xA,\xT,\xV,\xS} \draw[life] (\x,-0.36) -- (\x,\yCtr);

% heads
\pic[scale=0.8] at (\xO,0.72) {ov-server};
\node[head] at (\xO,0) {Orchestrator};
\pic[scale=0.8] at (\xA,0.72) {ov-robot};
\node[head] at (\xA,0) {MLE agent\\(ReAct loop)};
\pic[scale=0.8] at (\xT,0.72) {ov-tree};
\node[head] at (\xT,0) {Solution tree};
\pic[scale=0.8] at (\xV,0.72) {ov-shield};
\node[head] at (\xV,0) {Submission\\validator};
\begin{scope}
  \colorlet{iconc}{MainColor}
  \pic[scale=0.8] at (\xS,0.72) {ov-book};
\end{scope}
\node[head, draw=MainColor, fill=MainFill, text=MainColor] at (\xS,0) {Insight skills};

% activations
\draw[act] ({\xA-0.07},-1.45) rectangle ({\xA+0.07},-6.75);
\draw[act] ({\xO-0.07},-6.75) rectangle ({\xO+0.07},-7.95);

% setup
\draw[msg] (\xO,-1.45) -- ({\xA-0.07},-1.45);
\node[lbl] at ({(\xO+\xA-1.1)/2},-1.45) {task, public data,\\\textcolor{MainColor}{insights (playbook)},\\\textcolor{MainColor}{insight descriptions (skill)}};

% development loop
\draw[black!60, line width=0.5pt] ({\xA-1.0},-1.9) rectangle ({\xS+0.9},-6.3);
\node[anchor=north west, draw=black!60, fill=white, font=\sffamily\scriptsize\bfseries, inner sep=1.5pt] (loop) at ({\xA-1.0},-1.9) {loop};
\node[anchor=west, black!70, fill=black!5, inner sep=1pt] at ({\xA+0.15},-1.9) {[until final submission or budget exhausted]};

\SelfCall{\xA}{-2.4}
\node[anchor=west, align=left] at ({\xA+0.55},-2.55) {LLM call:\\reason and act};

\draw[msg] ({\xA+0.07},-3.2) -- (\xT,-3.2);
\node[lbl] at ({(\xA+\xT)/2},-3.2) {execute solution};
\draw[ret] (\xT,-3.65) -- ({\xA+0.07},-3.65);
\node[lbl] at ({(\xA+\xT)/2},-3.65) {new node, val.\ score};

\draw[msg] ({\xA+0.07},-4.3) -- (\xV,-4.3);
\node[lbl] at ({(\xT+\xV)/2},-4.3) {validate submission};
\draw[ret] (\xV,-4.75) -- ({\xA+0.07},-4.75);
\node[lbl] at ({(\xT+\xV)/2},-4.75) {valid / format errors};

\draw[msg, ins] ({\xA+0.07},-5.4) -- (\xS,-5.4);
\node[lbl, ins] at ({(\xV+\xS)/2},-5.4) {read skill};
\draw[ret, ins] (\xS,-5.85) -- ({\xA+0.07},-5.85);
\node[lbl, ins] at ({(\xV+\xS)/2},-5.85) {insight details};

% final grading
\draw[msg] ({\xA-0.07},-6.75) -- ({\xO+0.07},-6.75);
\node[lbl] at ({(\xO+\xA-1.1)/2},-6.75) {submit final node};

\SelfCall{\xO}{-7.45}
\node[anchor=west] at ({\xO+0.55},-7.6) {grading: attach private-test set, rerun solution, compute metrics};
% \SelfCall{\xO}{-8.1}
% \node[anchor=west] at ({\xO+0.55},-8.25) {rerun the selected solution from scratch};
% \SelfCall{\xO}{-8.75}
% \node[anchor=west] at ({\xO+0.55},-8.9) {grade the submission with the private-test labels};
\end{tikzpicture}}
\vspace{-4pt}
\caption{Workflow of the \AgenticMLESystem, with the insights highlighted.}\label{fig:architecture}
\vspace{-6pt}
\Description{A sequence diagram with five lifelines: orchestrator, MLE agent, solution tree, submission validator, and insight skills. The orchestrator starts the agent with the task, public data, the domain insight, and the skills or playbook. In a loop, the agent calls the LLM, executes solutions that become nodes of the solution tree, validates submissions, and reads skills in skill mode. After the agent submits a final node, the orchestrator replaces the public-test inputs with the private-test ones, reruns the selected solution, and grades the submission with the private-test labels.}
\end{figure*}

The assembled insights can be injected into any \AgenticMLESystem that accepts textual context.
In our study, we inject them into a ReAct-style \AgenticMLESystem that develops solutions in an isolated container with access to the public data of the \TargetTask, whose workflow is shown in Figure~\ref{fig:architecture}.
The system can inspect data, edit files, execute candidate solutions, and validate submissions.
Each execution is recorded as a node in a solution tree, so the system can branch from an earlier candidate when a modification fails, similar to AIDE~\cite{JiangETAL25AIDE}.
At the deadline, the selected solution is rerun from scratch on the private-test inputs and graded outside the agent loop.
Without insights, this system is the \Baseline of our evaluation (\S\ref{sec:eval}).

\subsubsection{Insight Injection Modes}
\label{sec:technique:injection}

\begin{wrapfigure}{r}{0.42\textwidth}
\vspace{-12pt}
\centering
\begin{tikzpicture}[
  x=1cm, y=1cm, font=\sffamily\scriptsize,
  dom/.style={circle, draw=black!40, fill=black!5, line width=0.6pt},
  src/.style={circle, fill=black!70, inner sep=0pt, minimum size=4pt},
  tgt/.style={star, star points=5, star point ratio=2.2, fill=MainColor, draw=white, inner sep=0pt,
    minimum size=11pt},
  insnode/.style={draw=#1Color, fill=#1Fill, line width=0.6pt, inner sep=0pt},
  flow/.style={-{Latex[length=1.5mm]}, line width=0.6pt, shorten >=1.5pt, shorten <=1pt},
  dflow/.style={flow, DomainColor, line width=0.9pt},
  cflow/.style={flow, CompetitionColor},
  iflow/.style={flow, ImprovementColor},
  feed/.style={flow, #1Color!55, densely dashed, line width=0.5pt},
  ann/.style={inner sep=1pt, align=center},
]
% domain of the target task
\node[dom, draw=DomainColor, fill=DomainFill!60, minimum size=3.0cm] (D) at (0,0) {};
\node[ann, DomainColor!80!black, anchor=north] at (-0.3,-1.52) {current domain};
\node[tgt] (t) at (0.55,0.15) {};
\node[src] (s1) at (-0.9,0.45) {};
\node[src] (s2) at (-1.05,-0.2) {};
\node[src] (s3) at (-0.85,-0.65) {};

% other domains
\node[dom, minimum size=1.45cm] (O1) at (3.4,0.85) {};
\node[dom, minimum size=1.45cm] (O2) at (3.4,-0.95) {};
\node[ann, black!60, anchor=south] at (3.4,1.6) {other domains};
\node[src] (o1) at (3.15,1.15) {};
\node[src] (o2) at (3.65,1.05) {};
\node[src] (o3) at (3.3,0.5) {};
\node[src] (o4) at (3.15,-0.55) {};
\node[src] (o5) at (3.7,-1.0) {};
\node[src] (o6) at (3.25,-1.35) {};

% domain insight: same-domain sources -> domain insight -> target task
\node[insnode=Domain, minimum size=7pt] (di) at (D.65) {};
\node[ann, DomainColor, anchor=south] at ([yshift=1pt]di.north) {domain insight};
\foreach \s in {s1,s2,s3} \draw[feed=Domain] (\s) -- (di);
\draw[dflow] (di) -- (t);

% competition insights: same-domain sources -> target task
\foreach \s in {s1,s2,s3} \draw[cflow] (\s) -- (t);

% improvement insights: sources of all domains -> improvement insights -> target task
\node[insnode=Improvement, minimum width=10pt, minimum height=7pt,
  double copy shadow={shadow xshift=1.5pt, shadow yshift=-1.5pt}] (ii) at (2.15,-0.75) {};
\node[ann, ImprovementColor, anchor=north] at ([xshift=-9pt, yshift=-2.5pt]ii.south) {improvement\\insights};
\draw[feed=Improvement] (s1) to[bend right=12] (ii);
\foreach \s in {s2,s3} \draw[feed=Improvement] (\s) -- (ii);
\foreach \s in {o1,o2,o3,o4,o5,o6} \draw[feed=Improvement] (\s) -- (ii);
\draw[iflow, line width=0.9pt] (ii) -- (t);

\node[ann, CompetitionColor!80!black, fill=DomainFill!60, inner sep=0.5pt, anchor=north] at (0.5,-0.3) {competition\\insights};

% legend
\node[src] at (-1.45,-2.25) {};
\node[ann, anchor=west] at (-1.35,-2.25) {source competition};
\node[tgt, minimum size=8pt] at (1.25,-2.25) {};
\node[ann, anchor=west] at (1.4,-2.25) {target task};
\end{tikzpicture}
\vspace{-4pt}
\caption{Scope of the insights injected for a \TargetTask.}\label{fig:insight-scope}
%  (star): the \DomainInsight and the \CompetitionInsights of its own domain, and the \ImprovementInsights assembled from all domains
\Description{A large circle represents the current domain and contains the target task, drawn as a star, and three \SourceCompetitions, drawn as dots. Two smaller circles represent other domains, each containing three \SourceCompetitions. Dashed arrows go from the \SourceCompetitions in the current domain to a green square labeled domain insight, and from the \SourceCompetitions of all domains to a red stack of rectangles labeled improvement insights; solid arrows go from the square, from the stack, and from each \SourceCompetition in the current domain (labeled competition insights) to the target task.}
\vspace{-8pt}
\end{wrapfigure}
For a \TargetTask, the system receives the \DomainInsight of the task's domain, the \CompetitionInsights of the \SourceCompetitions in the same domain, and all \ImprovementInsights, which are assembled from the \SourceCompetitions of every domain (Figure~\ref{fig:insight-scope}).
The two injection modes hold the insights, system, prompt, harness, task, and budget constant, and only change \emph{when} the \CompetitionInsights and \ImprovementInsights enter the context; the \DomainInsight is always in the initial context.
% The \DomainInsight of the \TargetTask's domain is placed in the initial context in both modes, since it provides a broad prior for the task.
% The \CompetitionInsights of the peer competitions in the same domain and the \ImprovementInsights are injected either as \Skills or as a \Playbook.
% In both modes, the insights of the \TargetTask itself are never available (\S\ref{sec:benchmark}).

\MyPara{\SkillModeCap}
Following the progressive-disclosure design of Agent Skills~\cite{Anthropic25AgentSkills}, the initial context lists only compact metadata (name and short description) of each available insight.
% The full insight texts and their evidence are mounted read-only in the container, and the system loads them on demand through dedicated read tools or ordinary file reads.
This mode preserves context for task observations and execution feedback, but its benefit depends on the system recognizing a relevant insight from its description and retrieving it.

\MyPara{\PlaybookModeCap}
Following the use of evolving playbooks as agent context~\cite{ZhangETAL26ACE}, the full texts of all available insights are rendered directly into the system prompt before the system begins.
% The skill directory and read tools are then removed so that the system cannot receive the same insight twice.
This mode removes the retrieval decision and makes every insight immediately available, at the cost of a longer context that may contain insights irrelevant to, or inconsistent with, the \TargetTask.

\subsubsection{Application to New Competitions}
\label{sec:technique:new}

The assembled insights are not tied to \Benchmark: a new competition from outside it can use them in the same way as a \TargetTask.
For one or a few new competitions, the only step needed is to determine the domain of each competition, either manually or by embedding its description as in Domain Clustering (\S\ref{sec:benchmark}) and assigning it to the closest cluster; the \DomainInsight and \CompetitionInsights of that domain and all \ImprovementInsights are then directly usable without further assembly.

Once a new competition has ended and its writeups and solutions are available, it can be added as a \SourceCompetition by an incremental update that reruns only the affected assembly tasks.
\Tool assembles the \CompetitionInsight of the new competition and re-synthesizes the \DomainInsight of its domain; if its writeups yield new \RawImprovementInsights, they are assigned to the existing clusters, and only the \ImprovementInsights of the clusters that receive them are re-synthesized.
Re-clustering is needed only when many competitions have been added, which may shift the cluster structure, or when a new competition belongs to no existing domain.
% Until a new domain is formed, a competition outside every domain can still use the \ImprovementInsights, which are assembled across domains. 
% Otherwise, the existing \CompetitionInsights and \RawImprovementInsights are reused, since each is assembled from a single competition or writeup. 
% and only the \DomainInsights and \ImprovementInsights of the changed clusters are re-synthesized.

\subsection{Running Example}
\label{sec:technique:example}

\begin{figure*}[t]
\centering
\begin{subfigure}{\textwidth}
\input{figs/example-insights}
\vspace{-2pt}
\caption{Excerpts of the three types of insights available for the task.}
% ; each insight has a front matter (name and description, in gray) followed by its content.}
\label{fig:example-insights}
\end{subfigure}

\smallskip
\begin{subfigure}{\textwidth}
\input{figs/example-solutions}
\caption{Excerpts of the final solutions without and with insights; numbered changes relate to (a).}\label{fig:example-solutions}
\end{subfigure}
\vspace{-8pt}
\caption{Running example on \ExampleTask.}
% ; ``[\ldots]'' and ``...'' mark omitted text and code.}
\label{fig:example}
\Description{Two parts. Part (a) shows three side-by-side insight files, each with a gray front matter holding its name and description, followed by its content. The Time Series Prediction domain insight has three items under the DATA heading, each stating a condition, an action, and a rationale; the highlighted actions are (1) to use target lags and rolling baselines anchored at least one horizon behind each target, (2) to normalize historical rates to the denominator available for the forecast period, and (3) to align each backtest value to its actual publication vintage. The GoDaddy Microbusiness Density Forecasting competition insight has three items under the DATA heading; the highlighted actions are (2) to normalize all historical rates to the denominator expected at forecast time and (3) to align each value to its real publication vintage, whereas the item on entities with tiny counts is not highlighted. The deployment-contract-temporal-backtesting improvement insight recommends gating every candidate with a deployment-contract backtest replayed across several forecast origins, with the highlighted rationale (3) of matching the as-of information boundary. Part (b) shows a side-by-side diff of two code excerpts with three hunks: features, training, and prediction through the time-series API. The solution without insights, with a validation MAE of 29.24 and a private-test MAE of 234.15, computes rolling target means shifted by only two hours, joins the client data of the same day, trains on the raw target, and at prediction time merges client data from the training period only and fills missing features with zeros. The solution with insights, with a validation MAE of 31.91 and a private-test MAE of 57.12, computes rolling means from target lags of at least 48 hours (change 1), divides the production target by the installed capacity in training and multiplies the predictions by it (change 2), and joins the client data as of two days earlier, both in training and from the data served by the API (change 3).}
\end{figure*}

We illustrate the assembly and usage of insights with the competition \ExampleTask, the \TargetTask of the Time Series Prediction domain.
The task asks for hourly forecasts of the electricity produced and consumed by Estonian customers with solar panels, given weather forecasts, energy prices, and client data such as the installed photovoltaic capacity, and is evaluated by mean absolute error (MAE).
Predictions are made through a time-series API that serves the test data day by day, in which the latest revealed targets and client data lag two days behind the prediction times.
The four \SourceCompetitions of the domain forecast COVID-19 cases, microbusiness density, and retail sales, and none of them involves energy.

\MyPara{Insights}
Figure~\ref{fig:example-insights} shows excerpts of the three types of insights available for this task.
% Although they are assembled from other tasks, several \InsightItems have triggers that hold for \ExampleTask.
Three aspects of the \DomainInsight have triggers that hold for \ExampleTask: \CircledNum{1} target lags anchored at least one horizon behind each target match the two-day delay of the revealed targets, \CircledNum{2} the denominator normalization matches the production that scales with the installed capacity, which grows over time, and \CircledNum{3} aligning external features to their publication vintage matches the two-day delay of the client data.
The \CompetitionInsight of GoDaddy Microbusiness Density Forecasting~\cite{godaddy-microbusiness-density-forecasting} contains the items from which \CircledNum{2} and \CircledNum{3} are generalized, whereas other items, such as the one on entities with tiny counts, have triggers that do not hold and can be skipped.
The \ImprovementInsight \Code{deployment-contract-temporal-backtesting} recommends accepting a candidate solution only if it improves in a backtest that replays the serving contract, i.e., the available history, the gap, and the horizon, across several forecast origins, which requires \CircledNum{3} matching the information available at each origin.
% \genai{In the six \FullSkill runs on this task, the \DomainInsight is always in the context, whereas the \Skills are rarely retrieved: one run reads \Code{deployment-contract-temporal-backtesting} before developing its first solution, and no run reads the \CompetitionInsight.}

\MyPara{Impact on the solutions}
Over six runs each, the \AgenticMLESystem reaches a private-test MAE of \UseMacro{N-ExampleBaselineMean}$\pm$ \UseMacro{N-ExampleBaselineMargin} without insights and \UseMacro{N-ExamplePlaybookMean}$\pm$ \UseMacro{N-ExamplePlaybookMargin} with insights (lower is better, since MAE measures the prediction error).
Figure~\ref{fig:example-solutions} compares the final solutions of a run without insights and a run with insights, in which all three insights are injected via the \PlaybookMode.
The solution without insights anchors its rolling target means only two hours before each prediction time and joins the client data of the same day, although the API reveals the targets and client data only two days later.
Its hold-out validation on the training data cannot detect this mismatch and reports an MAE of \UseMacro{N-ExampleRunBaselineValidation}; at prediction time, the rolling means and the capacity are missing (the client data served by the API is ignored), filled with zeros, and the private-test MAE rises to \UseMacro{N-ExampleRunBaselinePrivate}.
The solution with insights makes three changes that implement the highlighted insight parts: \CircledNum{1} it computes its rolling means only from target lags of at least 48 hours, \CircledNum{2} it learns the production per unit of installed capacity and rescales its predictions by the capacity, and \CircledNum{3} it joins the client data as of two days earlier, both in training and from the data served by the API.
% The trajectory does not reveal which insight prompted each change, since all of them are in the context.
Its validation MAE (\UseMacro{N-ExampleRunPlaybookValidation}) is worse but reflects the deployment setting, and its private-test MAE is \UseMacro{N-ExampleRunPlaybookPrivate}.

\section{Evaluation}
\label{sec:eval}

We answer the following research questions:

\begin{itemize}[topsep=3pt,itemsep=0.5ex,partopsep=0ex,parsep=0ex,leftmargin=*]
\item \textbf{RQ1}: How much do the assembled insights improve the solutions of an \AgenticMLESystem?
\item \textbf{RQ2}: How do the three types of insights contribute to the improvement?
\item \textbf{RQ3}: How do the two injection modes, \Skills and \Playbooks, compare in effectiveness and cost?
\item \textbf{RQ4}: Do the improvements generalize across backend LLMs?
\end{itemize}

\subsection{Experimental Setup}
\label{sec:eval:setup}

\MyPara{\TargetTasksCap and system}
We evaluate on the \UseMacro{N-EvalCompetitions} held-out \TargetTasks of \Benchmark (Table~\ref{tab:benchmark}), using the insights that \Tool assembled from the \UseMacro{N-CorpusCompetitions} \SourceCompetitions.
The base \AgenticMLESystem is backed by \DeepSeekModel.
Each run receives one NVIDIA RTX A6000 GPU (48G of RAM built with cuda\_11.5.r11.5), four AMD EPYC 7343 CPUs (16 cores @ 1.5GHz), a 12-hour development budget, and at most 250 agent steps.
% Table~\ref{tab:setup} summarizes the common setup.

% \begin{table}[t]
% \begin{small}
% \begin{center}
% \caption{Common experimental setup.}\label{tab:setup}
% \input{tables/table-setup}
% \end{center}
% \end{small}
% \end{table}

\begin{wraptable}{r}{0.5\textwidth}
\begin{footnotesize}
\vspace{-12pt}
\begin{center}
\caption{Insight injection modes of each condition.}\label{tab:modes}
\setlength{\tabcolsep}{4pt}
\begin{tabular}{@{}llll@{}}
\toprule
\UseMacro{TH-Mode} &
\UseMacro{TH-DomainInsight} &
\UseMacro{TH-CompetitionInsight} &
\UseMacro{TH-ImprovementInsight} \\
\midrule
\Baseline & -- & -- & -- \\
\FullSkill & playbook & skill & skill \\
\FullPlaybook & playbook & playbook & playbook \\
\midrule
% \DCSkill & playbook & skill & -- \\  % hidden for now
\DomainPlaybook & playbook & -- & -- \\
\CompetitionPlaybook & -- & playbook & -- \\
\ImprovementPlaybook & -- & -- & playbook \\
\bottomrule
\end{tabular}

\end{center}
\vspace{-8pt}
\end{footnotesize}
\end{wraptable}

\MyPara{Conditions}
Table~\ref{tab:modes} lists the six conditions in our experiments, combining different injection modes.
\Baseline receives no insight.
\FullSkill and \FullPlaybook inject all three types of insights, as \Skills and as a \Playbook, respectively (\S\ref{sec:technique:injection}).
Specially, \DomainInsights are always injected as \Playbook since they are short and most relevant.
The three ablation conditions, \DomainPlaybook, \CompetitionPlaybook, and \ImprovementPlaybook, each inject a single type of insight  (\DomainInsights, \CompetitionInsights, and \ImprovementInsights, respectively) as a \Playbook.
% DC-skill hidden for now (SHOW_DC_SKILL in automation/generate_eval_{results,figures}.py).
% \DCSkill injects \DomainInsights as a \Playbook and \CompetitionInsights as \Skills, without \ImprovementInsights.

\MyPara{Metrics}
The primary metric is the private-test score of the final solution, computed by the task's \MLEBench-compatible grader.
Because raw scores are not comparable across \Benchmark tasks, we min--max normalize the scores of each task on a fixed range, after reversing lower-is-better metrics, and compute the average improvement over the baseline for each mode across all tasks:
\begin{equation}
    \label{eq:Delta}
    \Delta=
    \frac{\Delta_m - \Delta_{Baseline}}{\Delta_{Baseline}} \times 100\%,
    \quad
    \Delta_m=
        \frac{1}{T}
        \sum_{t=1}^{T}\frac{1}{K}\sum_{k=1}^K N_{m,t,k},
    \quad
    N_{m,t,k}
    =
    \frac{x_{m,t,k}-L_t}{U_t-L_t},
\end{equation}
where \(\Delta_m\) and \(\Delta_{Baseline}\) denote the average normalized scores of mode \(m\) and \Baseline over \(T\) tasks (\(T=\UseMacro{N-EvalCompetitions}\)) and \(K\) repetitions (\(K=6\)), \(N_{m,t,k}\) and \(x_{m,t,k}\) are the normalized and raw scores of mode \(m\) on task \(t\) in the \(k\)-th run, and \(L_t\) and \(U_t\) are the worst and best scores over the runs of \Baseline, \FullSkill, and \FullPlaybook on task \(t\). The same range is used for every condition to put all conditions on one scale, so an ablation run outside it may be normalized to below 0 or above 1.

To measure cost, we record the input, reasoning, and output tokens consumed by the backend LLM and the wall-clock time of each run.
We convert the tokens to a monetary cost at the OpenRouter API prices of \DeepSeekModel: \$0.44 per million uncached input tokens, \$0.014 per million cached input tokens, and \$1.32 per million output tokens, where the output tokens include the reasoning tokens.
For each task and condition, we report the mean with a 95\% Student-$t$ confidence interval, and compare each condition with \Baseline using a two-sided Mann--Whitney $U$ test~\cite{Mann1947Utest} at $\alpha=0.05$.
Since each condition is compared with \Baseline on \UseMacro{N-EvalCompetitions} tasks, we also report the number of significant tasks after the Holm--Bonferroni correction~\cite{Holm79Holm} over the tasks of each condition.
Every condition has \UseMacro{N-BaselineRuns} runs, i.e., six runs per task.
% Which runs count when a condition/task has extra (failed or late) attempts is decided manually on
% the experimental server; do not discuss run selection in the paper.
In addition to the private-test score of the final solution, we analyze the validation scores that the system measures on its own validation split for every scored candidate solution during development, which show how the solution quality evolves over time.
% For each task, we flip the validation scores if needed so that higher is better and min--max scale them to $[0,100]$ over all runs of the compared conditions; 
The validation scores are normalized in the same way as private-test scores;
at each minute, we take the median over the runs of a task and average it over tasks.

\begin{table}[t]
\begin{small}
\begin{center}
\caption{Overall results.}\label{tab:results-main}
\setlength{\tabcolsep}{4pt}
\FitTable{\begin{tabular}{@{}lrrr@{\hspace{14pt}}rrr@{}}
\toprule
 & \UseMacro{TH-Baseline} & \UseMacro{TH-Skill} & \UseMacro{TH-Playbook} & \UseMacro{TH-DomainPlaybook} & \UseMacro{TH-CompetitionPlaybook} & \UseMacro{TH-ImprovementPlaybook} \\
\midrule
Normalized private-test score & \UseMacro{N-BaselineScore} & \UseMacro{N-SkillScore} & \UseMacro{N-PlaybookScore} & \UseMacro{N-DomainPlaybookScore} & \UseMacro{N-CompetitionPlaybookScore} & \UseMacro{N-ImprovementPlaybookScore} \\
\UseMacro{TH-ChangeFromBaseline} & -- & \UseMacro{N-SkillGain} & \UseMacro{N-PlaybookGain} & \UseMacro{N-DomainPlaybookGain} & \UseMacro{N-CompetitionPlaybookGain} & \UseMacro{N-ImprovementPlaybookGain} \\
Normalized validation score & \UseMacro{N-BaselineValidationScore} & \UseMacro{N-SkillValidationScore} & \UseMacro{N-PlaybookValidationScore} & \UseMacro{N-DomainPlaybookValidationScore} & \UseMacro{N-CompetitionPlaybookValidationScore} & \UseMacro{N-ImprovementPlaybookValidationScore} \\
\midrule
Per-task input tokens & \UseMacro{N-BaselineInputTokens} & \UseMacro{N-SkillInputTokens} & \UseMacro{N-PlaybookInputTokens} & \UseMacro{N-DomainPlaybookInputTokens} & \UseMacro{N-CompetitionPlaybookInputTokens} & \UseMacro{N-ImprovementPlaybookInputTokens} \\
Per-task reasoning tokens & \UseMacro{N-BaselineReasoningTokens} & \UseMacro{N-SkillReasoningTokens} & \UseMacro{N-PlaybookReasoningTokens} & \UseMacro{N-DomainPlaybookReasoningTokens} & \UseMacro{N-CompetitionPlaybookReasoningTokens} & \UseMacro{N-ImprovementPlaybookReasoningTokens} \\
Per-task output tokens & \UseMacro{N-BaselineOutputTokens} & \UseMacro{N-SkillOutputTokens} & \UseMacro{N-PlaybookOutputTokens} & \UseMacro{N-DomainPlaybookOutputTokens} & \UseMacro{N-CompetitionPlaybookOutputTokens} & \UseMacro{N-ImprovementPlaybookOutputTokens} \\
\UseMacro{TH-PerTaskTime} & \UseMacro{N-BaselineTimeMin} & \UseMacro{N-SkillTimeMin} & \UseMacro{N-PlaybookTimeMin} & \UseMacro{N-DomainPlaybookTimeMin} & \UseMacro{N-CompetitionPlaybookTimeMin} & \UseMacro{N-ImprovementPlaybookTimeMin} \\
\UseMacro{TH-TotalTime} & \UseMacro{N-BaselineGPUDay} & \UseMacro{N-SkillGPUDay} & \UseMacro{N-PlaybookGPUDay} & \UseMacro{N-DomainPlaybookGPUDay} & \UseMacro{N-CompetitionPlaybookGPUDay} & \UseMacro{N-ImprovementPlaybookGPUDay} \\
\midrule
\UseMacro{TH-PerTaskCost} & \UseMacro{N-BaselineCostAvg} & \UseMacro{N-SkillCostAvg} & \UseMacro{N-PlaybookCostAvg} & \UseMacro{N-DomainPlaybookCostAvg} & \UseMacro{N-CompetitionPlaybookCostAvg} & \UseMacro{N-ImprovementPlaybookCostAvg} \\
\UseMacro{TH-TotalCost} & \UseMacro{N-BaselineCostTotal} & \UseMacro{N-SkillCostTotal} & \UseMacro{N-PlaybookCostTotal} & \UseMacro{N-DomainPlaybookCostTotal} & \UseMacro{N-CompetitionPlaybookCostTotal} & \UseMacro{N-ImprovementPlaybookCostTotal} \\
\bottomrule
\end{tabular}
}
\end{center}
\end{small}
\end{table}

\subsection{RQ1: Effectiveness of Insights}
\label{sec:eval:rq1}

\begin{table}[t]
\begin{footnotesize}
\begin{center}
\caption{Per-task change of the normalized score from \Baseline (bold: significant).}\label{tab:results-per-task}
\setlength{\tabcolsep}{3pt}
\FitTable{%% Auto-generated by automation/generate_eval_results.py from LLM-Insight-Graph@fddb0b72:README.md (deepseek-v4-flash-0731, gpt-6.1-sol).
\begin{tabular}{@{}lrrrrr@{}}
\toprule
\UseMacro{TH-Competition} & \UseMacro{TH-Skill} & \UseMacro{TH-Playbook} & \UseMacro{TH-DomainPlaybook} & \UseMacro{TH-CompetitionPlaybook} & \UseMacro{TH-ImprovementPlaybook} \\
\midrule
Ariel Data Challenge 2025 & \textbf{186.6\%} & \textbf{193.8\%} & \textbf{154.4\%} & \textbf{180.0\%} & \textbf{232.2\%} \\
CAFA 5 Protein Function & \textbf{116.2\%} & \textbf{198.4\%} & \textbf{162.0\%} & 118.3\% & 91.6\% \\
CSIRO Biomass & \textbf{176.6\%} & \textbf{193.5\%} & \textbf{151.5\%} & \textbf{123.4\%} & \textbf{214.6\%} \\
Energy Behavior of Prosumers & \textbf{149.4\%} & \textbf{148.9\%} & \textbf{106.7\%} & \textbf{132.9\%} & \textbf{112.5\%} \\
Image Matching 2025 & \textbf{680.8\%} & \textbf{613.4\%} & 174.2\% & \textbf{430.5\%} & 169.9\% \\
Jigsaw Community Rules & \textbf{236.9\%} & \textbf{215.4\%} & 155.3\% & \textbf{153.1\%} & \textbf{197.4\%} \\
March Mania 2025 & \textbf{110.6\%} & \textbf{107.5\%} & \textbf{79.8\%} & \textbf{110.7\%} & \textbf{76.3\%} \\
Mitsui Commodity Prediction & \textbf{229.2\%} & \textbf{282.8\%} & \textbf{248.0\%} & 39.4\% & \textbf{248.4\%} \\
NFL Player Contact & \textbf{209.6\%} & \textbf{243.0\%} & \textbf{195.4\%} & 186.3\% & \textbf{237.1\%} \\
POG Series S01E01 & 7.2\% & -24.2\% & 3.8\% & -1.4\% & -12.7\% \\
PhysioNet ECG Digitization & \textbf{118.0\%} & \textbf{67.9\%} & \textbf{83.2\%} & \textbf{150.5\%} & \textbf{116.3\%} \\
RSNA Aneurysm Detection & \textbf{84.7\%} & \textbf{100.5\%} & \textbf{78.6\%} & \textbf{82.0\%} & 63.6\% \\
\midrule
Tasks improved & \UseMacro{N-SkillTasksImproved} & \UseMacro{N-PlaybookTasksImproved} & \UseMacro{N-DomainPlaybookTasksImproved} & \UseMacro{N-CompetitionPlaybookTasksImproved} & \UseMacro{N-ImprovementPlaybookTasksImproved} \\
Tasks improved significantly & \UseMacro{N-SkillTasksSignificant} & \UseMacro{N-PlaybookTasksSignificant} & \UseMacro{N-DomainPlaybookTasksSignificant} & \UseMacro{N-CompetitionPlaybookTasksSignificant} & \UseMacro{N-ImprovementPlaybookTasksSignificant} \\
\bottomrule
\end{tabular}
}
\end{center}
\end{footnotesize}
\end{table}

\begin{figure}[t]
\centering
\includegraphics[width=\linewidth]{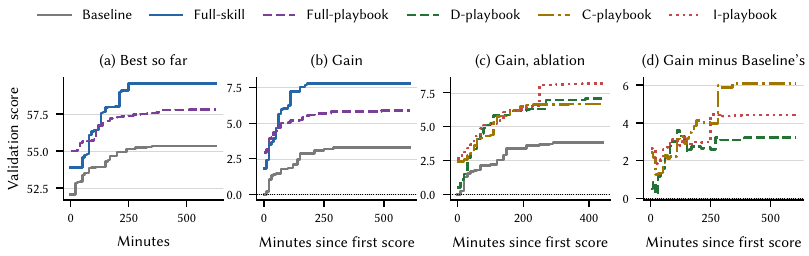}
\vspace{-8pt}
\caption{Validation scores during development; gains are over the first solution of \Baseline.}\label{fig:eval-curves}
\vspace{-6pt}
\Description{Four line charts over development minutes. (a) The best validation score found so far: Full-skill ends highest at about 60, Full-playbook at about 58, and Baseline at about 55; Full-skill and Full-playbook are above Baseline from the start. (b) The gain over Baseline's first solution: Full-skill and Full-playbook start at about 2 and 3 points and end at about 8 and 6 points, and Baseline starts at 0 and ends at about 3 points. (c) All three single-type conditions start above Baseline and end with gains of about 7 to 8 points, compared with about 4 points for Baseline. (d) The difference to Baseline is positive throughout for all three single-type conditions, at about 2 to 3 points in the first 100 minutes; C-playbook ends highest at about 6 points, I-playbook at about 4, and D-playbook at about 3.}
\end{figure}

As shown in Tables~\ref{tab:results-main} and~\ref{tab:results-per-task}, injecting the assembled insights improves the normalized score from \UseMacro{N-BaselineScore} for \Baseline to \UseMacro{N-SkillScore} for \FullSkill (\UseMacro{N-SkillGain}) and \UseMacro{N-PlaybookScore} for \FullPlaybook (\UseMacro{N-PlaybookGain}).
The improvement is consistent across tasks: \FullSkill improves the mean score on \UseMacro{N-SkillTasksImproved} of the \UseMacro{N-EvalCompetitions} tasks, significantly so on \UseMacro{N-SkillTasksSignificant} of them (\UseMacro{N-SkillTasksSignificantHolm} after the Holm--Bonferroni correction), with POG Series S01E01 as the only task where the improvement is not significant; \FullPlaybook improves on \UseMacro{N-PlaybookTasksImproved} tasks, significantly on \UseMacro{N-PlaybookTasksSignificant} (\UseMacro{N-PlaybookTasksSignificantHolm} after the correction).
The largest gains occur on tasks, such as Image Matching 2025 and Energy Behavior of Prosumers, where the validation design and task-specific modeling choices documented in peer competitions are consequential.

Figure~\ref{fig:eval-curves}(a, b) shows that the insights help both in drafting the initial solution and in improving it: the first scored solutions of \FullSkill and \FullPlaybook are \UseMacro{N-CurveSkillGainFirst} and \UseMacro{N-CurvePlaybookGainFirst} points above those of \Baseline; they reach the final best score of \Baseline after \UseMacro{N-CurveSkillReachBaselineMin} and \UseMacro{N-CurvePlaybookReachBaselineMin} minutes, and end with gains of \UseMacro{N-CurveSkillGainFinal} and \UseMacro{N-CurvePlaybookGainFinal} points over the first solution of \Baseline, compared with \UseMacro{N-CurveBaselineGainFinal} points for \Baseline itself.

\begin{RQAnswer}{RQ1}
The assembled insights improve the solution quality of the \AgenticMLESystem by \UseMacro{N-SkillGain} (\FullSkill) and \UseMacro{N-PlaybookGain} (\FullPlaybook) in normalized score, with \FullSkill improving on \UseMacro{N-SkillTasksImproved} of the \UseMacro{N-EvalCompetitions} \TargetTasks; the system with insights starts from better solutions and improves them faster.
\end{RQAnswer}

\subsection{RQ2: Contribution of Insight Types}
\label{sec:eval:rq2}

We compare \FullPlaybook with the single-type ablations (last three columns of Table~\ref{tab:results-main}).

\begin{figure}[t]
\centering
% the generated panel titles (top 12pt) are clipped and replaced by the subcaptions
\captionsetup[subfigure]{skip=2pt}
\begin{subfigure}{0.48\linewidth}
\centering
\includegraphics[trim=0 0 139.26 12, clip]{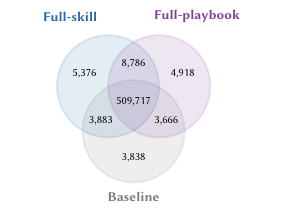}
\caption{\Baseline, \FullSkill, and \FullPlaybook}
\end{subfigure}\hfill
\begin{subfigure}{0.48\linewidth}
\centering
\includegraphics[trim=137 0 0 12, clip]{figs/eval-venn.pdf}
\caption{\Baseline and the single-type ablations}
\end{subfigure}
\vspace{-8pt}
\caption{Private-test items correctly predicted only by the enclosing conditions.}\label{fig:eval-venn}
% \genai{, i.e., the unique units graded in the private-test set of a task} (e.g., protein--term pairs or player pairs)
\Description{Two Venn diagrams. (a) Baseline, Full-skill, and Full-playbook: 509,717 items are predicted correctly by all three; 3,838 only by Baseline, 5,376 only by Full-skill, 4,918 only by Full-playbook, and 8,786 by Full-skill and Full-playbook but not Baseline. (b) Baseline and the three single-type ablations: 509,377 items are predicted correctly by all four; 1,184 only by Baseline, 1,238 only by D-playbook, 12,345 only by C-playbook, and 4,566 only by I-playbook.}
\vspace{-6pt}
\end{figure}

Each type of insight alone improves the normalized score over \Baseline (\UseMacro{N-BaselineScore}): to \UseMacro{N-DomainPlaybookScore} with \DomainInsights (\UseMacro{N-DomainPlaybookGain}), \UseMacro{N-CompetitionPlaybookScore} with \CompetitionInsights (\UseMacro{N-CompetitionPlaybookGain}), and \UseMacro{N-ImprovementPlaybookScore} with \ImprovementInsights (\UseMacro{N-ImprovementPlaybookGain}).
All three remain below \FullPlaybook (\UseMacro{N-PlaybookScore}).
Hence, the three types are complementary rather than redundant, and their benefits accumulate when combined.
The accumulation is not additive, but the combination is more robust: the best single type varies across tasks (e.g., \ImprovementPlaybook on Ariel Data Challenge 2025 and CSIRO Biomass, and \CompetitionPlaybook on PhysioNet ECG Digitization), whereas \FullPlaybook has a higher mean score than all three single-type conditions on \UseMacro{N-PlaybookAboveSingleTypesTasks} of the \UseMacro{N-EvalCompetitions} tasks.
They are also less consistent: their improvement is significant on only \UseMacro{N-DomainPlaybookTasksSignificant}, \UseMacro{N-CompetitionPlaybookTasksSignificant}, and \UseMacro{N-ImprovementPlaybookTasksSignificant} tasks (\UseMacro{N-DomainPlaybookTasksSignificantHolm}, \UseMacro{N-CompetitionPlaybookTasksSignificantHolm}, and \UseMacro{N-ImprovementPlaybookTasksSignificantHolm} after the correction), compared with \UseMacro{N-PlaybookTasksSignificant} (\UseMacro{N-PlaybookTasksSignificantHolm}) for \FullPlaybook (Table~\ref{tab:results-per-task}).

Figure~\ref{fig:eval-curves}(c, d) shows the validation scores of the ablation modes during development.
All three single-type modes are ahead of \Baseline from their first scored solution on, by similar margins in the first 100 minutes (\UseMacro{N-CurveDomainPlaybookEffectAtHundred}, \UseMacro{N-CurveCompetitionPlaybookEffectAtHundred}, and \UseMacro{N-CurveImprovementPlaybookEffectAtHundred} points); afterwards, \CompetitionPlaybook pulls ahead and ends \UseMacro{N-CurveCompetitionPlaybookEffectFinal} points above \Baseline, compared with \UseMacro{N-CurveImprovementPlaybookEffectFinal} and \UseMacro{N-CurveDomainPlaybookEffectFinal} points for \ImprovementPlaybook and \DomainPlaybook.
On the five tasks graded on per-item correctness (CAFA 5, Jigsaw, March Mania, NFL, and RSNA), we also compare the private-test items that each condition predicts correctly, i.e., that at least four of its six runs predict correctly.
Each type also yields private-test items that no other condition predicts correctly (Figure~\ref{fig:eval-venn}(b)): \UseMacro{N-VennCompetitionPlaybookOnly} with \CompetitionInsights, \UseMacro{N-VennImprovementPlaybookOnly} with \ImprovementInsights, and \UseMacro{N-VennDomainPlaybookOnly} with \DomainInsights, compared with \UseMacro{N-VennAblationBaselineOnly} for \Baseline.

\begin{RQAnswer}{RQ2}
The three types of insights are complementary: each type alone improves the score, but significantly on fewer tasks and less than the combination of all three types; \CompetitionInsights contribute the most to the late-stage improvements and to the predictions that no other condition gets right.
\end{RQAnswer}

\subsection{RQ3: Comparison of Injection Modes}
\label{sec:eval:rq3}

Overall, the two injection modes reach similar normalized scores (\UseMacro{N-SkillScore} for \FullSkill and \UseMacro{N-PlaybookScore} for \FullPlaybook in Table~\ref{tab:results-main}), but the better mode varies across tasks.
\FullPlaybook has the higher mean score on \UseMacro{N-PlaybookAheadTasks} tasks and \FullSkill on \UseMacro{N-SkillAheadTasks}; notably, \FullPlaybook falls below \Baseline on POG Series S01E01 (though not significantly), whereas \FullSkill stays above it.
A \Playbook guarantees that every insight is seen but can steer the system toward insights that do not fit the task, whereas \Skills let the system skip irrelevant insights but depend on it retrieving the relevant ones.
% Comparing the two modes directly, their difference is significant only on POG Series S01E01 (Mann--Whitney $U$ test, $p<0.05$ without correction), and a Wilcoxon signed-rank test over the per-task mean normalized scores finds no significant difference ($p=\UseMacro{N-SkillVsPlaybookWilcoxonP}$).
Although \FullSkill reaches a similar score to \FullPlaybook, its trajectories show that the system rarely retrieves the full text of a \Skill: only \UseMacro{N-SkillReadsReadRuns} of its \UseMacro{N-SkillReadsRuns} runs call \Code{read\_skill} at all, \UseMacro{N-SkillReadsReadCompetitionRuns} of them to read a \CompetitionInsight (typically that of the most similar \SourceCompetition, e.g., Image Matching Challenge 2024~\cite{image-matching-challenge-2024} for Image Matching 2025) and \UseMacro{N-SkillReadsReadImprovementRuns} to read an \ImprovementInsight.
The improvement of \FullSkill therefore stems mostly from what every run receives in its initial context rather than from the retrieved texts: the \DomainInsight, with which \DomainPlaybook alone reaches \UseMacro{N-DomainPlaybookScore}, and the metadata of the \Skills, whose descriptions state when each insight applies, e.g., the problem that each \ImprovementInsight resolves.
The two modes also lead to different correct predictions (Figure~\ref{fig:eval-venn}(a)): \UseMacro{N-VennSkillOnly} private-test items are predicted correctly only by \FullSkill and \UseMacro{N-VennPlaybookOnly} only by \FullPlaybook, compared with \UseMacro{N-VennBaselineOnly} only by \Baseline.

The two modes differ in cost.
Relative to \Baseline, \FullSkill increases the per-task input tokens by only \UseMacro{N-SkillInputTokensChange}, whereas \FullPlaybook increases them by \UseMacro{N-PlaybookInputTokensChange}, because the \Playbook is carried in the context of every model call; the output tokens and time are similar (Table~\ref{tab:results-main}).
The per-task cost is only \UseMacro{N-BaselineCostAvg}, \UseMacro{N-SkillCostAvg}, and \UseMacro{N-PlaybookCostAvg}, respectively, as most input tokens hit the prompt cache.

\begin{RQAnswer}{RQ3}
The better injection mode varies across tasks; overall, \Skills are more cost-effective, reaching a similar score to \Playbooks with \UseMacro{N-SkillInputTokensChange} instead of \UseMacro{N-PlaybookInputTokensChange} additional input tokens.
\end{RQAnswer}

\subsection{RQ4: Generalizability across LLMs}
\label{sec:eval:rq4}

We repeat \Baseline, \FullSkill, and \FullPlaybook using \GPTModel, a proprietary LLM perceived to be stronger than \DeepSeekModel, on two of the \TargetTasks: Image Matching 2025, where the insights bring the largest gain (\UseMacro{N-DeepSeekIMCSkillGain} with \FullSkill, \UseMacro{N-DeepSeekIMCPlaybookGain} with \FullPlaybook), and POG Series S01E01, the only task with no significant improvement (\UseMacro{N-DeepSeekPOGSkillGain} with \FullSkill, \UseMacro{N-DeepSeekPOGPlaybookGain} with \FullPlaybook).
We use the same insights and six runs per task.

\begin{table}[t]
\begin{small}
\begin{center}
\caption{Private-test scores (mean$\pm$95\% CI) with two backend LLMs (bold: significant).}\label{tab:results-llms}
\FitTable{%% Auto-generated by automation/generate_eval_results.py from LLM-Insight-Graph@fddb0b72:README.md (deepseek-v4-flash-0731, gpt-6.1-sol).
\begin{tabular}{@{}l@{\hspace{3pt}}lrrr@{}}
\toprule
\UseMacro{TH-Competition} & \UseMacro{TH-LLM} & \UseMacro{TH-Baseline} & \UseMacro{TH-Skill} & \UseMacro{TH-Playbook} \\
\midrule
\multicolumn{5}{@{}l}{\textit{Image Matching 2025} (mAA$\uparrow$)} \\
 & \DeepSeekModel & 21.1\tiny{$\pm$1.8} & \textbf{35.2\tiny{$\pm$6.7}} & \textbf{33.8\tiny{$\pm$5.4}} \\
 & \GPTModel & 42.8\tiny{$\pm$1.6} & \textbf{52.0\tiny{$\pm$9.3}} & \textbf{49.9\tiny{$\pm$10.5}} \\
\midrule
\multicolumn{5}{@{}l}{\textit{POG Series S01E01} (MAE$\downarrow$)} \\
 & \DeepSeekModel & 0.0255\tiny{$\pm$0.0009} & 0.0249\tiny{$\pm$0.0004} & 0.0273\tiny{$\pm$0.0031} \\
 & \GPTModel & 0.0261\tiny{$\pm$0.0014} & 0.0251\tiny{$\pm$0.0006} & 0.0253\tiny{$\pm$0.0005} \\
\midrule
\multicolumn{5}{@{}l}{\textit{Overall}} \\
\UseMacro{TH-ChangeFromBaseline} & \DeepSeekModel & -- & 61.8\% & 27.5\% \\
 & \GPTModel & -- & 80.3\% & 61.9\% \\
\UseMacro{TH-TotalTime} & \DeepSeekModel & 1.7 & 2.1 & 2.5 \\
 & \GPTModel & 3.6 & 3.7 & 3.9 \\
\UseMacro{TH-TotalCost} & \DeepSeekModel & \$22 & \$21 & \$22 \\
 & \GPTModel & \$153 & \$128 & \$160 \\
\bottomrule
\end{tabular}
}
\end{center}
\end{small}
\end{table}

As shown in Table~\ref{tab:results-llms}, the insights are not specific to one backend LLM: on the normalized scale over both tasks, they also improve \GPTModel by \UseMacro{N-GPTSkillGain} with \FullSkill and \UseMacro{N-GPTPlaybookGain} with \FullPlaybook.
\GPTModel is a stronger backend than \DeepSeekModel on Image Matching 2025: its \Baseline already reaches an mAA of \UseMacro{N-GPTIMCBaselineScore}, close to the gold-medal threshold of the human leaderboard (\UseMacro{N-IMCGoldThreshold}).
Although the improvement is smaller than with \DeepSeekModel, the insights still significantly improve the normalized score by \UseMacro{N-GPTIMCSkillGain} with \FullSkill and \UseMacro{N-GPTIMCPlaybookGain} with \FullPlaybook, lifting the mean mAA beyond the gold-medal threshold (\UseMacro{N-GPTIMCSkillScore} and \UseMacro{N-GPTIMCPlaybookScore}).
On POG Series S01E01, where the insights do not help \DeepSeekModel, they improve the normalized score of \GPTModel by \UseMacro{N-GPTPOGSkillGain} with \FullSkill and \UseMacro{N-GPTPOGPlaybookGain} with \FullPlaybook, although not significantly over six runs; this suggests that a stronger backend LLM can make better use of the insights.
Still, no condition with either backend LLM reaches the median of the human leaderboard on this task (MAE \UseMacro{N-POGMedianThreshold}), so the tasks of \Benchmark remain challenging for future \AgenticMLESystems.
Unlike \DeepSeekModel, \GPTModel actively retrieves \Skills: \UseMacro{N-GPTSkillReadsReadRuns} of its \UseMacro{N-GPTSkillReadsRuns} \FullSkill runs call \Code{read\_skill} (\UseMacro{N-GPTSkillReadsReadImprovementRuns} for \ImprovementInsights and \UseMacro{N-GPTSkillReadsReadCompetitionRuns} for \CompetitionInsights), which may explain why \FullSkill beats \FullPlaybook with \GPTModel on both tasks.

The two backend LLMs also develop solutions differently: on Image Matching 2025, \GPTModel creates its first solution node earlier (median step \UseMacro{N-GPTIMCFirstNodeStep} vs.\ \UseMacro{N-DeepSeekIMCFirstNodeStep}), uses most of the 12-hour budget (\UseMacro{N-GPTIMCBaselineTimeMin}--\UseMacro{N-GPTIMCSkillTimeMin} minutes), and tries \UseMacro{N-GPTIMCBaselineNodes}--\UseMacro{N-GPTIMCPlaybookNodes} solution nodes per run, whereas \DeepSeekModel usually submits its final solution after \UseMacro{N-DeepSeekIMCSkillTimeMin}--\UseMacro{N-DeepSeekIMCPlaybookTimeMin} minutes with \UseMacro{N-DeepSeekIMCSkillNodes}--\UseMacro{N-DeepSeekIMCBaselineNodes} nodes.

\begin{RQAnswer}{RQ4}
The improvements from insights generalize to \GPTModel, a stronger proprietary LLM from a different model family: the insights improve its normalized score by \UseMacro{N-GPTSkillGain} (\FullSkill) and \UseMacro{N-GPTPlaybookGain} (\FullPlaybook), lifting it beyond the gold-medal threshold on Image Matching 2025.
Still, some MLE tasks remain challenging to LLM agents: on POG Series S01E01, neither \DeepSeekModel nor \GPTModel reaches the median of the human leaderboard, with or without insights.
\end{RQAnswer}

\section{Limitations}
\label{sec:limitations}

\MyPara{Task scope}
\Benchmark is derived from Kaggle competitions, which provide realistic end-to-end \MLE tasks but do not represent every production setting.
Kaggle emphasizes predictive performance under a fixed metric and submission interface, whereas deployed machine learning systems also require maintainability, monitoring, fairness, and privacy.
% Moreover, the \UseMacro{N-EvalCompetitions} \TargetTasks cover heterogeneous domains, but each domain contributes only one \TargetTask.

\MyPara{Domain construction}
The domains of \Benchmark are induced by clustering competition descriptions.
Other representations or clustering choices could change which competitions are considered closely related and thus contribute insights to a \TargetTask.

\MyPara{Insight quality}
The insights are assembled by an LLM-based agent.
To reduce the risk of hallucination and misrepresentation, \Tool must read every required source in full and back each insight with an evidence record, which links the insight to the human-written passages it is based on and makes it traceable.
% Consolidating \UseMacro{N-ImprovementInsights} \RawImprovementInsights into \UseMacro{N-SynthesizedImprovements} \ImprovementInsights may also merge strategies whose applicability conditions differ.
% \genai{\Fix{XXX: Conduct a blinded expert audit of a stratified sample of insights and report agreement and error categories.}}

\MyPara{Leakage beyond the benchmark}
\Benchmark excludes the materials of every \TargetTask from the insight sources, but it cannot control what the backend LLMs have learned during pre-training.
Because Kaggle writeups and notebooks are public, a backend LLM may have seen solutions of the \TargetTasks; this affects all conditions equally, including \Baseline, and therefore may inflate absolute scores.
Selecting the most recent eligible competition of each domain as the \TargetTask reduces, but does not eliminate, this risk.

\MyPara{Models and costs}
The main results use one backend LLM for the \AgenticMLESystem and one LLM for \Tool, so they may depend on these models' knowledge, tool-use behavior, and long-context capabilities; RQ4 studies the former dependency.
The 12-hour, GPU-backed budget also limits accessibility and may favor approaches that spend more compute.
Assembling the insights from all \UseMacro{N-CorpusCompetitions} \SourceCompetitions with \AssemblyModel cost \$\UseMacro{N-AssemblyCost} USD (\$\UseMacro{N-AssemblyCostAll} USD including runs repeated during development), a one-time cost that is amortized over all tasks of a domain.
% ; we did not evaluate settings with smaller development budgets.
% \Fix{XXX: Report the performance under at least one lower-budget setting.}
% the ranges are the min/max over the conditions of Table~\ref{tab:results-main}; update if the numbers change
The experiments are also costly: with \DeepSeekModel, each condition takes \UseMacro{N-BaselineGPUDay}--\UseMacro{N-CompetitionPlaybookGPUDay} GPU days and \UseMacro{N-BaselineCostTotal}--\UseMacro{N-DomainPlaybookCostTotal} in API costs, and with \GPTModel, the API costs of a condition on all \UseMacro{N-EvalCompetitions} \TargetTasks would reach \UseMacro{N-GPTProjectedCostMax}, extrapolated from up to \UseMacro{N-GPTLLMTasksPlaybookCostTotal} on two tasks.
% \genai{Each combination of backend LLM and condition is also costly to evaluate: with \DeepSeekModel, the \UseMacro{N-BaselineRuns} runs of one condition take \UseMacro{N-BaselineGPUDay}--\UseMacro{N-CompetitionPlaybookGPUDay} GPU days and \UseMacro{N-BaselineCostTotal}--\UseMacro{N-DomainPlaybookCostTotal} in API costs, and with \GPTModel, the \UseMacro{N-GPTBaselineRuns} runs of one condition on only two tasks already take up to \UseMacro{N-GPTLLMTasksPlaybookGPUDay} GPU days and \UseMacro{N-GPTLLMTasksPlaybookCostTotal}; this limits the number of backend LLMs, conditions, and \TargetTasks that we could evaluate.}
% Finally, 
% proprietary LLMs evolve quickly: the insights were assembled with \AssemblyModel, whereas RQ4 already uses the newer \GPTModel, and model versions may be updated or retired before others can rerun our experiments.
Running \Tool with a different backend LLM may yield different findings, which our budget precludes evaluating; to facilitate reproduction and such evaluations by others, we open-source our workflow and insights.
%  the workflow remains reproducible, and we release the assembled insights so that the evaluation can be repeated without reassembling them.

\section{Related Work}
\label{sec:related}

\MyPara{\AgenticMLECap}
MLAgentBench~\cite{HuangETAL24MLAgentBench} evaluates LLM agents that design, run, and revise ML experiments, and \MLEBench~\cite{ChanETAL25MLEBench} evaluates end-to-end solutions of Kaggle competitions against private-test labels.
The systems on the \MLEBench leaderboard mostly advance the search over candidate solutions~\cite{JiangETAL25AIDE,LiuETAL25MLMaster,ToledoETAL25AIRADojo,LiETAL25FMAgent,WanETAL25LoongFlow,YangETAL25RDAgent} and, increasingly, the use of insights from other tasks (\S\ref{sec:background:insights}).
Systems that use insights couple a particular source of insights with a particular agent scaffold, and evaluate the combination as a whole on \MLEBench.
Our work is complementary: rather than proposing another scaffold, we hold the \AgenticMLESystem fixed and study how the assembled insights, their types, and their injection modes affect it on a benchmark whose \SourceCompetitions are fixed and disjoint from the \TargetTasks.

\MyPara{Program synthesis for ML code}
Before LLM agents, program synthesis targeted ML code by searching a restricted space of programs.
BLAZE~\cite{WangETAL18BLAZE} synthesizes tensor transformations in a MATLAB-like DSL from input--output examples, and TF-Coder~\cite{ShiETAL22TFCoder} enumerates expressions over TensorFlow operations, prioritized by learned models; INTENT~\cite{ZhouETAL22INTENT} lets users decompose hard tensor transformations, and HySynth~\cite{BarkeETAL24HySynth} guides the search with a grammar learned from LLM completions.
Nam et al.~\cite{NamETAL22APICentric} predict the sequence of PyTorch and TensorFlow API calls that transforms the inputs into the outputs to guide enumerative search.
At a coarser granularity, AutoML-Zero~\cite{RealETAL20AutoMLZero} evolves complete ML algorithms from basic mathematical operations, NEAR~\cite{ShahETAL20NEAR} and dPADS~\cite{CuiAndZhu21dPADS} search for differentiable programs in a DSL, and Marinescu et al.~\cite{MarinescuETAL21CFGPipelines} search ML pipelines defined by a context-free grammar.
These techniques synthesize short programs or pipelines within a predefined search space, whereas an \AgenticMLESystem writes an end-to-end solution in a general-purpose language; the insights assembled by \Tool play a role similar to their learned priors, steering the system toward decisions that worked on related tasks.

\MyPara{Experience, memory, and context for LLM agents}
Reflexion~\cite{ShinnETAL23Reflexion} stores verbal feedback from earlier trials of the same task, ExpeL~\cite{ZhaoETAL24ExpeL} extracts natural-language lessons from successful and failed trajectories and retrieves them for new tasks, and Agent Workflow Memory~\cite{WangETAL25AgentWorkflowMemory} induces reusable workflows from past trajectories.
ACE~\cite{ZhangETAL26ACE} treats the agent's context as an evolving playbook that is incrementally refined from execution feedback, and Agent Skills~\cite{Anthropic25AgentSkills} package instructions and resources that the agent loads on demand.
These approaches mainly learn from the agent's own trajectories.
\Tool instead assembles insights from human-written solutions and writeups of peer tasks, organizes them into domain, competition, and improvement insights, and compares the skill and playbook modes of injecting the same insights.

\MyPara{Mining Kaggle artifacts}
% \MetaKaggle~\cite{KaggleMetaKaggle} and Meta Kaggle Code~\cite{KaggleMetaKaggleCode} publish metadata and notebook source for public Kaggle competitions.
DistilKaggle~\cite{MostafaviETAL24DistilKaggle} curates Kaggle notebooks together with code-quality measures, and Jin et al.~\cite{JinETAL26Modernization} mine Kaggle notebooks and use \MLEBench-style grading to study their reproducibility under modern execution environments.
We use Kaggle descriptions, writeups, and top solutions as sources from which to assemble insights for solving unseen tasks.

\MyPara{Data contamination in benchmarks}
Contamination of evaluation data threatens the validity of LLM benchmarks, and LiveCodeBench~\cite{JainETAL25LiveCodeBench} mitigates it by continuously collecting problems released after the models' training cutoffs.
For \AgenticMLESystems, the relevant contamination also comes from insight sources available at solving time (\S\ref{sec:background:leakage}); \Benchmark controls these sources by construction, using only \NonDescendants of the \TargetTasks as sources.

\section{Conclusion}
\label{sec:conclusion}

We studied how insights assembled from closely-related \MLE tasks affect an \AgenticMLESystem.
We collected \Benchmark, whose held-out split of \UseMacro{N-BenchmarkCompetitions} Kaggle competitions in \UseMacro{N-Domains} domains keeps \TargetTask information out of the insights, and designed \Tool, an agent that assembles evidence-backed \DomainInsights, \CompetitionInsights, and \ImprovementInsights from the \SourceCompetitions and injects them as \Skills or a \Playbook.
The assembled insights improve the normalized private-test score by \UseMacro{N-SkillGain} as \Skills and \UseMacro{N-PlaybookGain} as a \Playbook; the three types of insights are complementary, \Skills are more cost-effective, and the improvements generalize.
These findings suggest treating insight assembly as an explicit component of \AgenticMLESystems whose sources are controlled in evaluations.
% We envision our work as a foundation for future research on more effective insight assembly and \AgenticMLESystems, towards solving challenging MLE tasks such as POG Series S01E01.

% \section*{Data Availability} % does not count towards page limit

% Our artifact~\cite{Artifact} is publicly available at \ArtifactURL.
% It includes \Benchmark, the implementation of \Tool and the base \AgenticMLESystem, the assembled insights, and the scripts for reproducing our experiments.
% Instead of redistributing the Kaggle data, the artifact provides scripts that download each competition's data from Kaggle, subject to the competition's rules.

%% Acknowledgments
%% \begin{acks}                            %% acks environment is optional
                                        %% contents suppressed with 'anonymous'
  %% Commands \grantsponsor{<sponsorID>}{<name>}{<url>} and
  %% \grantnum[<url>]{<sponsorID>}{<number>} should be used to
  %% acknowledge financial support and will be used by metadata
  %% extraction tools.
  %% This material is based upon work supported by the
  %% \grantsponsor{GS100000001}{National Science
  %%   Foundation}{http://dx.doi.org/10.13039/100000001} under Grant
  %% No.~\grantnum{GS100000001}{nnnnnnn} and Grant
  %% No.~\grantnum{GS100000001}{mmmmmmm}.  Any opinions, findings, and
  %% conclusions or recommendations expressed in this material are those
  %% of the author and do not necessarily reflect the views of the
  %% National Science Foundation.

%% \end{acks}

\bibliography{bib}

%% Appendix
%% \section{Appendix}

%% Text of appendix \ldots

\end{document}